\documentclass[prb,aps,twocolumn,superscriptaddress]{revtex4-1}
\usepackage{graphicx,color}
\usepackage{amsthm}
\usepackage{amsfonts}
\usepackage{enumerate}
\usepackage{latexsym}
\usepackage{amsmath}
\usepackage{amssymb}
\usepackage[colorlinks=true,citecolor=blue,linkcolor=blue]{hyperref}

\emergencystretch=\maxdimen
\include{MyCommand}

\usepackage{soul} 

\begin{document}

\title{Determinant quantum Monte Carlo study of
exhaustion in the periodic Anderson model}

\author{Lufeng Zhang}
\affiliation{Department of Physics, Beijing Normal University, Beijing
100875, China\\}
\author{Tianxing Ma}
\email{txma@bnu.edu.cn}
\affiliation{Department of Physics, Beijing Normal University, Beijing
100875, China\\}
\author{Natanael C. Costa}
\affiliation{International School for Advanced Studies (SISSA),
Via Bonomea 265, 34136, Trieste, Italy}
\affiliation{Instituto de F\'isica, Universidade Federal do Rio de
Janeiro, Caixa Postal 68528, 21941-972 Rio de Janeiro, RJ, Brazil}
\author{Raimundo R. dos Santos}
\affiliation{Instituto de F\'isica, Universidade Federal do Rio de
Janeiro, Caixa Postal 68528, 21941-972 Rio de Janeiro, RJ, Brazil}
\author{Richard T. Scalettar}
\affiliation{Physics Department, University of California, Davis, California 95616, USA}

\begin{abstract}
The Kondo and periodic Anderson models describe many of
the qualitative features of local moments coupled to a conduction band,
and thereby the physics of materials such as the heavy fermions.  In
particular, when the exchange coupling $J$ or hybridization $V$ between
the moments and the electrons of the metallic band is large, singlets
form, quenching the magnetism.  In the opposite, small $J$ or $V$,
limit, the moments survive and the conduction electrons mediate an
effective interaction which can trigger long-range, often
antiferromagnetic order. In the case of the Kondo model, where
the moments are described by local spins, Nozi\`eres considered the
possibility that the available conduction electrons within the Kondo
temperature of the Fermi surface would be insufficient in number to
accomplish the screening.  Much effort in the literature has been
devoted to the study of the temperature scales in the resulting ``exhaustion" problem and how the ``coherence temperature" where a heavy
Fermi liquid forms is related to the Kondo temperature. In this paper,
we study a version of the periodic Anderson model in which some of the conduction electrons
are removed in a way which avoids the fermion sign problem and hence
allows low-temperature quantum Monte Carlo simulations which can
access both singlet formation and magnetic ordering temperature scales.
We are then able to focus on a somewhat different aspect of exhaustion
physics than previously considered: the effect of dilution on the
critical $V$ for the singlet-antiferromagnetic transition.
\end{abstract}

\date{\today}

\maketitle

\section{Introduction} \label{Sec:Intro}

A fundamental property of the description of a local magnetic moment
embedded in a sea of conduction electrons provided by the Kondo model
(KM) and the single impurity Anderson model (SIAM)\cite{anderson61} is
the screening of the moment through the formation of a Kondo singlet, a
phenomenon which occurs below a characteristic Kondo temperature $T_{\rm
K}$.  This singlet formation is accompanied by the appearance of a
narrow resonant state at the Fermi energy and a large electronic
effective mass, enabling these models to provide a qualitative picture
of heavy fermion physics --the enhancement of specific heat and
magnetic susceptibility.\cite{stewart84,nozieres85,lee86}  Certain
features of this problem are amenable to exact analytic solution, {\it
e.g.},~via the Bethe {\it ansatz}.\cite{andrei83,schlottmann87}

The periodic Anderson model (PAM) extends the single impurity problem to
the dense limit, \textit{i.e.}, to a lattice of magnetic moments, raising
the possibility of the emergence of magnetic ordered states.  This may
occur due to an indirect coupling between local moments mediated by the
conduction-electron polarization (oscillations of the spin density),
which is known as the Ruderman-Kittel-Kasuya-Yosida (RKKY)
interaction.\cite{ruderman54,kasuya56,yosida57} The Fermi wavevector
$k_{\rm F}$ of the conduction electrons determines the oscillation
wavelength between moments separated by distance $R$,
$ J_{\rm RKKY}(R) \sim \big(k_{\rm F} \, {\rm cos}(2 k_{\rm F} R)\big)/ R^3$.
Thus the density of conduction electrons $n_c$, via $k_{\rm F}$, plays a
crucial role in the magnetic ordering
pattern.\cite{Xavier04,Peters15,costa17,Igoshev17,Zhong19}
Due to its importance to heavy fermion physics, the competition between
singlet formation and the magnetic ordering has been investigated
through many different methods, from
analytical\cite{vidhyadhiraja04,hewson93,tsunetsugu97,rice85,rice86} to
numerical.\cite{vekic95,jarrell95,rozenberg95,georges96,tahvildarzadeh97,vidhyadhiraja00,
pruschke00,capponi01,benlagra11,wu15,aulbach15,hu17,Schafer18} In
particular, quantum Monte Carlo (QMC) simulations\cite{vekic95,hu17} have
provided evidence of the existence of a quantum phase transition from a
staggered AF phase to a spin-liquid state in the
two-dimensional PAM at half filling.

This competition is strongly affected by the electronic density: as
$n_c$ is reduced, fewer conduction electrons are available to screen the
local moments. Indeed, Nozi\`eres introduced the idea of ``exhaustion''
to describe the increased difficulty in singlet formation.  Even when,
naively, $n_c$ is large, only conduction electrons within $k_{\rm B}T$
of the Fermi surface are available for screening. Thus, Nozi\`eres
suggested that the singlet formation would occur at an energy scale
called the ``coherence temperature'' $T_{\rm coh}$, much lower than the
Kondo temperature of the SIAM.  In this picture, a particular functional
form $T_{\rm coh} \sim N(E_{\rm F}) \, T_{\rm K}^2/ N_{\rm imp}$, where
$N(E_{F})$ is the density of states at the Fermi energy and $N_{\rm
imp}$ is the number of local moments, reflects the availability of only
those conduction electrons within $T_{\rm K}$ of the Fermi surface.

Considerable numerical
effort\cite{vidhyadhiraja00,meyer00,hewson93,ono91} has gone into
evaluating $T_{\rm coh}$ and its relation to $T_{\rm K}$, specifically
on the validity of Nozi\`eres' original suggestion $T_{\rm coh} \sim
T_{\rm K}^2$.  The situation is potentially complex for a number of
reasons.  First, the singlets formed at this scale could be rather
different from those envisioned in the simpler SIAM where a single
$f$ moment is screened by conduction electrons.  Instead, below $T_{\rm
coh}$, a much more complex tangle of spin correlations might emerge in
which $f$ electrons also screen each other, {\it i.e.},~singlets between
$f$ electrons develop.  Second, for the PAM, there are additional energy
scales associated with $f$ electron charge fluctuations.  In this case,
it has been found\cite{vidhyadhiraja00} that the detailed relation
$T_{\rm coh} \sim N(E_{\rm F}) \, T_{\rm K}^2/ ( \, \alpha(U_f,V) \,
N_{\rm imp} \,)$ is also affected by the scales of the on-site $U_f$ and
interband hopping $V$ energies, as opposed to a simple counting of  the
relative numbers of conduction and local electrons.
Despite the great experimental and theoretical effort\cite{Nakatsuji04,Curro04,Yang08,Yang08b,Shirer12,wirth16,Yang17,Jiang14,Jiang17,Costa18},
the exact relation between these two energy scales ($T_{\rm K}$ and $T_{\rm coh}$) is still an open question.

\begin{figure}[tbp]
\includegraphics[scale=0.36]{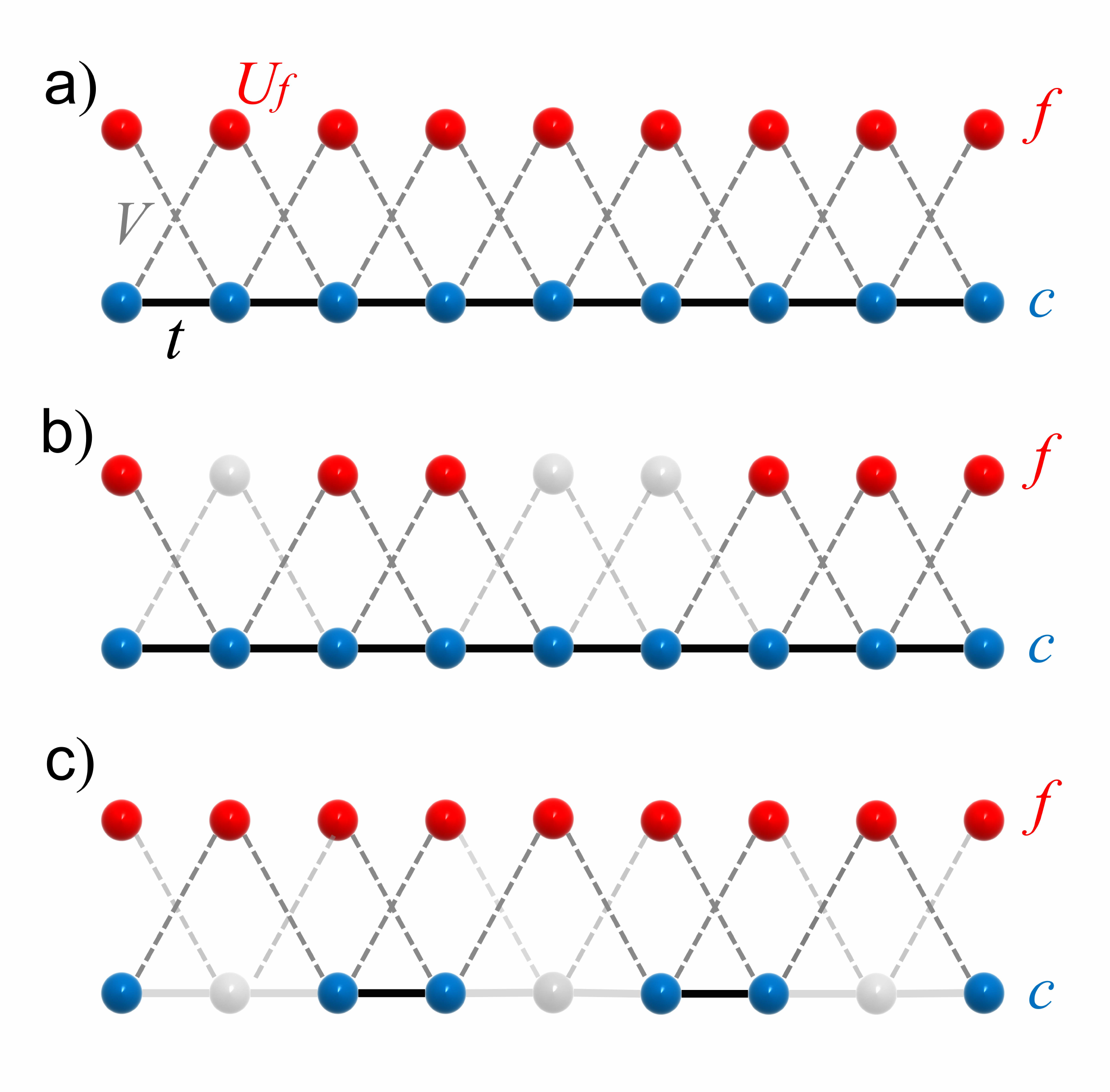}
\caption{One-dimensional representation of the geometry of our
Hamiltonian.  (a) The undiluted case in which
all conduction ($c$) and local ($f$) orbitals are present.
Bold horizontal lines are associated with the conduction
hopping $t$, and dashed diagonal lines with the
conduction-local electron hybridization $V$.
Both connect near-neighbor sites.
(b) Several of the $f$ orbitals, and their
associated hybridizations $V$ to the conduction orbitals, removed.
(c) Analogously, several of the conduction
orbitals removed.
This is the case most relevant to a study of exhaustion.}
\label{Fig:Figure1}
\end{figure}

Finally, it has been suggested\cite{nozieres98,meyer00} that the physics
of exhaustion might be fundamentally different depending on the strength
of the coupling between the conduction electrons and local moments.  For
large couplings, the singlets are local and dilution of conduction
electrons leaves behind well-defined local ``bachelor spins'' which must
then find a way to form singlets.  For small couplings, the screening is
largely collective in the first place, even before conduction-electron
dilution.  This would suggest that the nature of exhaustion differs markedly
at small and large $V$, in line with a more complex relation between
$T_{\rm coh}$ and $T_{\rm K}$ described in
Ref.\,\onlinecite{vidhyadhiraja00}.

In this work, our main interest is to investigate how the dilution of
conduction electrons affects the response of magnetic quantities in the
PAM.  In particular, and differently from previous
work,\cite{Kaul07,Watanabe10,Burdin18} we are interested in determining
the evolution of the quantum critical point (QCP) when the number of
conduction electrons differs from the localized ones, that is, $n_{c}
\neq n_{f}$.  In addition, we do not rely on Pauli blocking, {\it i.e.}
the restriction of conduction electron excitations to a temperature
window around $E_{\rm F}$.  Instead, we directly remove orbitals and
their associated electrons in order to introduce depletion.  In so
doing, we can examine not just Nozi\`ere's original exhaustion limit,
when the ratio $p=n_c/n_f < 1$, but the opposite case, $p=n_c/n_f > 1$,
as well.  This problem is investigated with the aid of an exact
numerical approach, namely the determinant quantum Monte Carlo (DQMC) method,\cite{blankenbecler81,hirsch85,white89a,assaad02,dosSantos03b,gubernatis16} by
introducing a model which allows us to control the ratio $p\equiv
n_c/n_f$ without running into the sign problem.\cite{loh90,troyer05}
The description of the model and the methodology are presented in the
next section.  Our results are shown in Sec. \ref{Sec:Results},
while our main conclusions are summarized in
Sec. \ref{Sec:Conclusions}.

\section{Model and numerical method}\label{Sec:Model}

Our work is focused on the PAM, whose Hamiltonian in real space reads
\begin{align}\label{Eq:Hamiltonian}
\hat H=-&\sum_{\langle i,j\rangle \sigma}
t_{ij}\big( \,
c^{\dagger}_{i\sigma}
c^{\phantom{\dagger}}_{j\sigma}
+
H.c.
\, \big)
-\sum_{\langle i,j\rangle\sigma}
V_{ij}\big( \,
c^{\dagger}_{i\sigma}
f^{\phantom{\dagger}}_{j\sigma}
+
H.c.
\, \big)
\nonumber \\
+ U_f &\sum_{{\bf i}} \big( \, n^f_{i\uparrow} -\frac{1}{2}\, \big)
\big( \, n^f_{i\downarrow} -\frac{1}{2} \, \big)
- \mu \sum_{{\bf i}}  n_{{\bf i}}
\,,
\end{align}
\vspace{0.15in}
Here,$c^{\dagger}_{i\sigma} \, (c^{\phantom{\dagger}}_{i\sigma})$ and $f^{\dagger}_{i\sigma} \,
(f^{\phantom{\dagger}}_{i\sigma})$ are the creation (annihilation)
operators of conduction and localized electrons, respectively, in the standard second quantization formalism.  Similarly, $n^c_{{\bf
i}\sigma}$ and $n^f_{{\bf i}\sigma}$ are site-number operators for $c$
and $f$ electrons, with $n_{\bf i} = \sum_{\sigma} (n^f_{{\bf i}\sigma}
+n^c_{{\bf i}\sigma})$ being the total occupation on site ${\bf i}$. $t_{ij}$ denotes the nearest-neighbor (NN) hopping between $c$ electrons, and $V_{ij}$ represents the nonlocal hybridization between $f$-orbitals and its NN conduction sites. In the $f$-orbital dilution case, when the $f$ electron is removed from site ${\bf i}$, $V_{ij}^{f\rightarrow c}=0\ (j\in\langle i,j\rangle)$ as Fig. \ref{Fig:Figure1}(b) shows.  On the other hand, if the $c$ electron on site ${\bf i}$ gets diluted, we set $t_{ij}=0, V_{ij}^{c\rightarrow f}=0\ (j\in\langle i,j\rangle)$ as the exhaustion case plotted in Fig. \ref{Fig:Figure1}(c).
Figure \ref{Fig:Figure1}(a) illustrates the nonlocal nature of the
hybridization for the undiluted geometry;
here, for simplicity, just the
one-dimensional analog is shown.
The local character of $f$ sites is
denoted by the momentum independence of the $f$ level, $\epsilon_f$, and
by the strong Coulomb repulsion $U_{f}$ for doubly occupied orbitals.
We set $t=1$ as the energy scale and explore the ``symmetric limit,'' $\mu=\epsilon_f=0$, for which the density of the $c$ and $f$ electrons is
half filled, $\langle n^{c}_{{i}\sigma} \rangle = \langle n^{f}_{{i}\sigma} \rangle = \frac{1}{2}$, a property which
holds for all $t, U_f, V,$ and temperatures $T$, due to the
particle-hole symmetry (PHS) of a bipartite lattice with NN hopping
terms.

At this point, we should mention that on-site (local) hybridizations are
more commonly studied rather than those with NN couplings, such as the doping effect of the on-site hybridization PAM.\cite{wei2017doping}  Then, it is
worth emphasizing the differences between both cases beyond their
dispersive character.  While the former leads to a charge gap for the
noninteracting limit ($U_f=0$) at half filling, the latter is a metal
for any hybridization strength.  As a consequence, the dispersive
$V_{\bf k}$ is more appropriate to describe a metallic
system.\cite{huscroft99,held00}
Further, in the case of on-site hybridization and $U_f\neq 0$, there is
evidence of conduction-electron localization when $f$ orbitals are
removed, accompanied by an enhancement of spin-spin correlations around
the unpaired noninteracting sites, breaking singlets and leading to a
magnetic ground state even at large
$V$.\cite{Titvinidze14,Titvinidze15,Aulbach15a,benali16,Costa18a} This effect may
overestimate the magnetic response, in particular, the value of the
critical hybridization for a given interaction strength, $V_{c}(U_{f})$.
By contrast, such effects are strongly attenuated in the dispersive
case since unpaired sites are less likely in a more connected lattice,
as in the case of nonlocal hybridization.  In view of this, the latter
seems more relevant to study the evolution of the critical point in
diluted systems.

The properties of the model are investigated
using the determinant quantum Monte Carlo method,\cite{blankenbecler81,hirsch85,white89a} which allows for an exact solution (to within statistical sampling errors) of the PAM Hamiltonian on finite-size  lattices.
Here we present highlights of the method, the details of which can be found
in a number of reviews; see, \textit{e.g.}, Refs.\,\onlinecite{gubernatis16,assaad02,dosSantos03b}.
The underlying step is the
construction of a path integral for the partition function ${\cal Z}$ by
discretizing the inverse temperature $\beta = L \Delta \tau$, and
breaking the full imaginary-time evolution operator $e^{-\beta \hat H}$
into incremental pieces $e^{-\Delta \tau \hat H}$.  This allows for the
use of the Trotter approximation,\cite{trotter59,suzuki76,fye86}
$e^{-\Delta \tau \hat H} \sim e^{-\Delta \tau \hat K} e^{-\Delta \tau
\hat {\cal U}},$ which isolates the interaction term $\hat {\cal U}$,
containing $U_f$, from the quadratic kinetic-energy pieces containing
$t, \mu,$ and $V$.

The interacting $\hat {\cal U}$ term is decoupled in a quadratic form by
performing a discrete Hubbard-Stratonovich (HS) transformation, with the
inclusion of auxiliary fields $S({\bf i},\tau)$ in both real and imaginary
coordinates, that are coupled to the spin of electrons.
Therefore, the path integral consists entirely
of quadratic forms and the fermionic trace can be
evaluated, resulting in a trace over HS fields of a product of
determinants,
${\rm det} {\cal M}_{\uparrow}(\{S({\bf i},\tau\}) \, {\rm
det} {\cal M}_{\downarrow}(\{S({\bf i},\tau)\}) $, of matrices whose
dimension is the number of spatial sites of the lattice.
The trace over $S({\bf i},\tau)$ is carried out by sampling them through
conventional Monte Carlo methods.
Here, in addition to the usual single moves, we also perform global moves,\cite{scalettar91}
which improves the ergodicity of the system.

Although the DQMC method is exact, it suffers from the infamous minus-sign
problem,\cite{loh90,troyer05} which arises from  the possibility of the
product ${\rm det} {\cal M}_{\uparrow}(\{S({\bf i},\tau)\} \, {\rm det}
{\cal M}_{\downarrow}(\{S({\bf i},\tau)\}$ being negative for certain
field configurations, corresponding to a negative density matrix.  The
sign problem is worse at low temperatures, large lattice sizes, or
strong interactions, and its dependence with the electron filling or
geometries is quite nontrivial.\cite{Mondaini12,iglovikov15} However,
this problem is absent for systems with PHS since it implies
constraints over the two determinants, leading to a positive total sign.
Notice that our Hamiltonian of Eq. \eqref{Eq:Hamiltonian} is
particle-hole symmetric at half filling, and hence the sign problem is
absent throughout this work.

As mentioned
earlier, the system is diluted through
the direct removal of orbitals and their associated electrons.
In fact, the more obvious way to reduce conducting electrons would be by simultaneously
setting $\mu <0$ and $\epsilon^f < 0$ since it preserves
the number of $f$ electrons, but lowers the $c$ occupancy.  However,
this leads to a severe sign problem and the energy scales for singlet
formation and antiferromagnetic (AF) order are no longer accessible.  By contrast, our
approach preserves PHS, which only requires that the hopping and
hybridization should be between NN sites at half filling, hence avoiding
any sign problem. Figures \ref{Fig:Figure1}(b) and
\ref{Fig:Figure1}(c) illustrate our dilution procedure for localized
($p=n_c/n_f >1$) and conduction ($p=n_c/n_f <1$) orbitals, respectively.
We should also note that  depletion affects the respective neighborhoods in different ways.
With the concentration of $c$ sites being $p=n_c/n_f < 1$, the probability that an $f$ site is connected to $0\leq m\leq 4$ active $c$ sites is
\begin{equation}
P_m= \frac{4!}{m!(4-m)!}\,p^m(1-p)^{4-m};
\label{eq:Pm}
\end{equation}
see the discussion of Figs.\,\ref{Fig:Figure8} and \ref{Fig:Figure9}.
In our measurements, for example of the structure factor below,
spatial configurations with isolated $f$ electrons ($m=0$) are not
included, since these sites contribute a ``trivial'' Curie-law free
moment $\chi \sim 1/T$.
Similarly, with the concentration of $f$ sites being $q=n_f/n_c < 1$,
the probability that a $c$ site is connected to $0\leq m\leq 4$ active
$f$ sites is also given by Eq. \eqref{eq:Pm}, but with $p$ being
replaced by $q$.

We investigate the magnetic properties by performing measurements of
spin-spin correlation functions for $f$ orbitals, and their Fourier
transform, the spin structure factor,
\begin{align}
{\cal S}^{\rm ff}(\pi,\pi) \equiv \frac{1}{N_{\rm f}} \sum_{\bf i,j}
\langle \, S^{\rm f}_{\bf i} \cdot S^{\rm f}_{\bf j}  \, \rangle
(-1)^{{\bf i}+{\bf j}},
\label{eq:Sff}
\end{align}
where $N_\mathrm{f}$ is the number of $f$ sites connected to at least
one $c$ site.  Here, we define the fermionic spin operators as
$ \vec S^{\rm f}_{\bf i} =
\big( \, f^{\dagger}_{{\bf i}\uparrow}
\, , f^{\dagger}_{{\bf i}\downarrow} \big)
\,\, \vec \sigma \,\,
\big( \, f^{\phantom{\dagger}}_{{\bf i}\uparrow}
\, , f^{\phantom{\dagger}}_{{\bf i}\downarrow} \big)^T \, ,
$
with  $\vec \sigma$ being the Pauli spin matrices; a similar expression
applies to $\vec S^c_{\bf i}$.  The phase factor
$(-1)^{{\bf i}+{\bf j}}$ takes opposite signs on the two sublattices,
corresponding to a staggered pattern.

\begin{figure}
\includegraphics[scale=0.42]{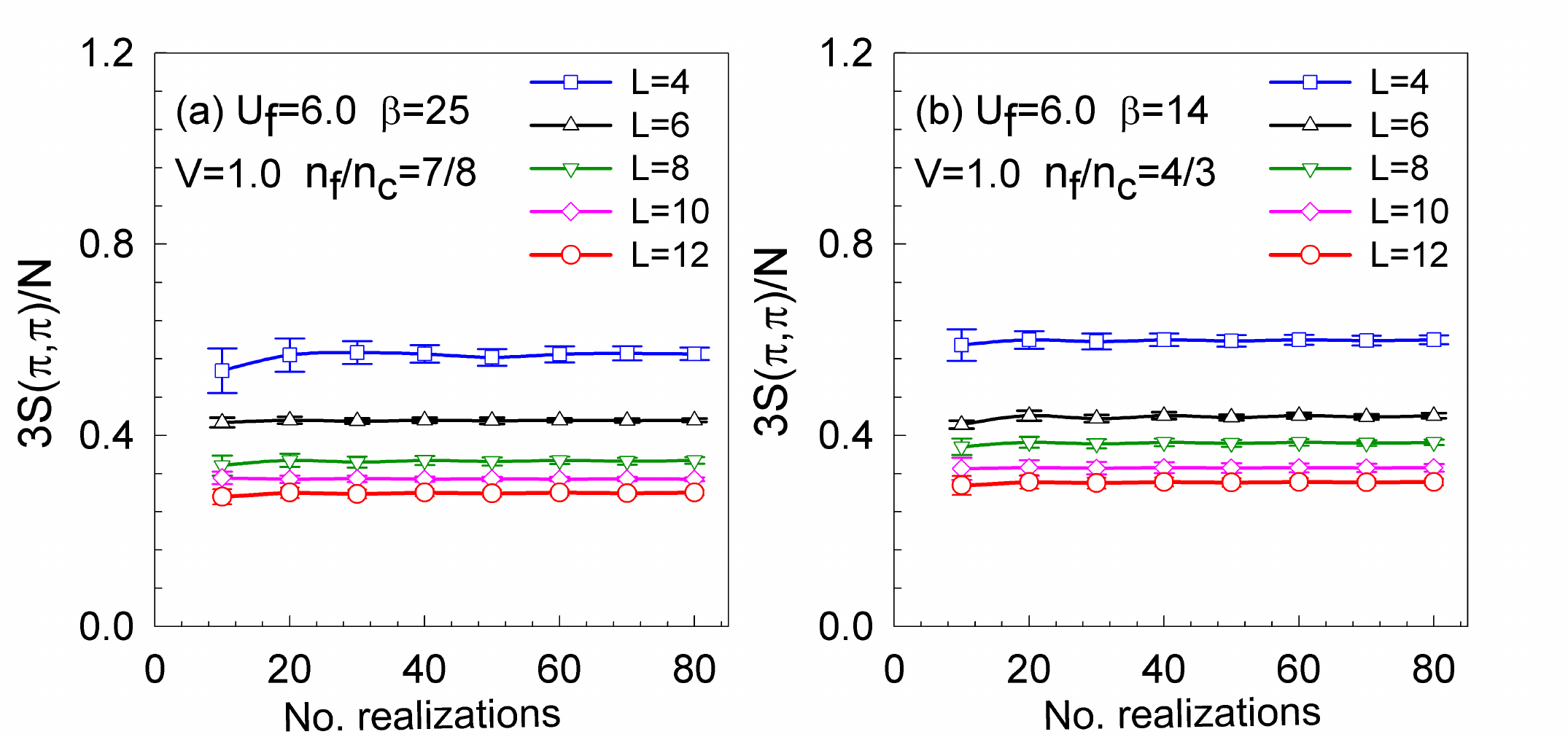}
\caption{(Color online) AF spin structure factor $S(\pi,\pi)$ computed on the $L=4,6,8,10,12$ lattices. At (a) $n_{f}/n_{c}=7/8$ and (b) $n_{f}/n_{c}=4/3$, in which the $f$ and $c$ electrons are diluted, and the data reported are obtained from different numbers of disorder realizations within the statistical errors.}
\label{Fig:Figure2}
\end{figure}

For singlet formation, we examine a correlator function,
\begin{align}
C^{fc}_{\bf i} \equiv \vec S^{\rm f}_{\bf i}  \cdot
\sideset{}{'}\sum_{{\bf j} \in {\cal N}({\bf i})} \vec S^{\rm c}_{\bf j} \, ,
\label{eq:Cfd}
\end{align}
with the sum being over sites ${\bf j}$ in the neighborhood ${\cal
N}({\bf i})$ of ${\bf i}$.  The prime on the sum emphasizes that in the
$c$-diluted case, some $f$ orbitals have less than four $c$ neighbors;
see Eq. \eqref{eq:Pm}.  By contrast, in the case of $f$ dilution, every
surviving $f$ site necessarily has four neighboring $c$ orbitals.
The data reported here are obtained from lattice sizes up to $L=12$,
with averaging over 20 different disorder realizations. In general the requisite number of realizations in simulations with disorder must be determined empirically and is a complex interplay between self-averaging on sufficiently large lattices, the strength of
the disorder, and the location in the phase diagram. We show the results in Fig. \ref{Fig:Figure2}. For any dilution case, the averaged $S(\pi,\pi)$ are consistent regardless of the number of realizations. These plots justify the use of 20 realizations in our work.
The error bars
shown in the following results reflect both statistical and disorder
sampling fluctuations.
\begin{figure}[tbp]
\includegraphics[scale=0.4]{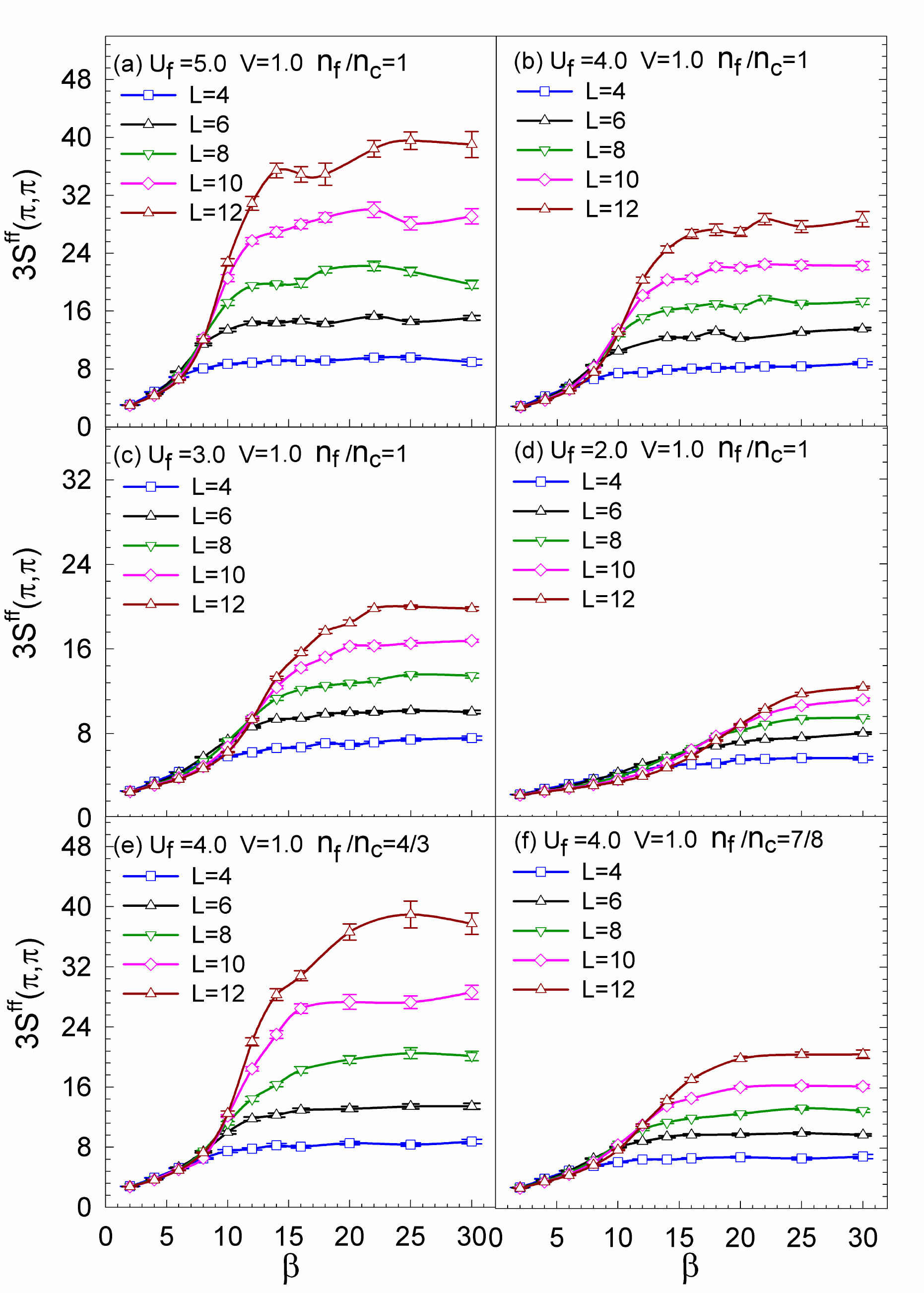}
\caption{
($a$)-($d$) Evolution of the AF structure factor, in the absence of
any dilution, with increasing $\beta$
for $V=1$ and different lattice sizes.
Results for $U_f=5,4,3,2$.
As $U_f$ decreases from $U_f=5$, larger $\beta$ values are required
for $S(\pi,\pi)$ to converge to the ground state limit.
The growth of $S(\pi,\pi)$ with lattice size suggests that
there may be long range AF order for this value of
$fd$ hybridization for all the $U_f$
shown.  ($e$),($f$) Analogous results
for $n_f \neq n_c$ at $U_f=4$.
}
\label{Fig:Figure3}
\end{figure}

\noindent
\section{Results and discussion} \label{Sec:Results}

Let us first discuss the undiluted PAM  [Fig. \ref{Fig:Figure1}(a)], with which the diluted case should be compared.
One should notice that due to the Mermin-Wagner theorem,\cite{Mermin66} long-range order is expected to occur only at $T=0$.
Therefore, as the temperature is lowered, the correlation length $\xi$ associated with spin correlations grows, but it is limited by the finite size of the system.
Figures \ref{Fig:Figure3}($a$)--\ref{Fig:Figure3}($d$) illustrate this for
the AF structure factor, plotted as a function of the inverse temperature, $\beta=1/T$, for different lattice sizes $L$ and $U_f$, fixing $V=1$.
At high temperatures (small $\beta$), $S^{\rm ff}(\pi,\pi)$ is independent
of lattice size, due to the short-range character of the spin correlations.
As $\beta$ increases, $\xi$ grows, ultimately reaching the
linear lattice size $L$, so that the structure factor increases and stabilizes at a finite value.
The growth of $S^{\rm ff}(\pi,\pi)$ with $L$
at low temperatures suggests the existence of long-range order, which should be verified through scaling arguments, as discussed below.
Figures \ref{Fig:Figure3}($e$)--\ref{Fig:Figure3}($f$) show that the behavior for the diluted cases is similar, irrespective of $n_f$ being larger or smaller than $n_c$.

\begin{figure}[tbp]
\includegraphics[scale=0.4]{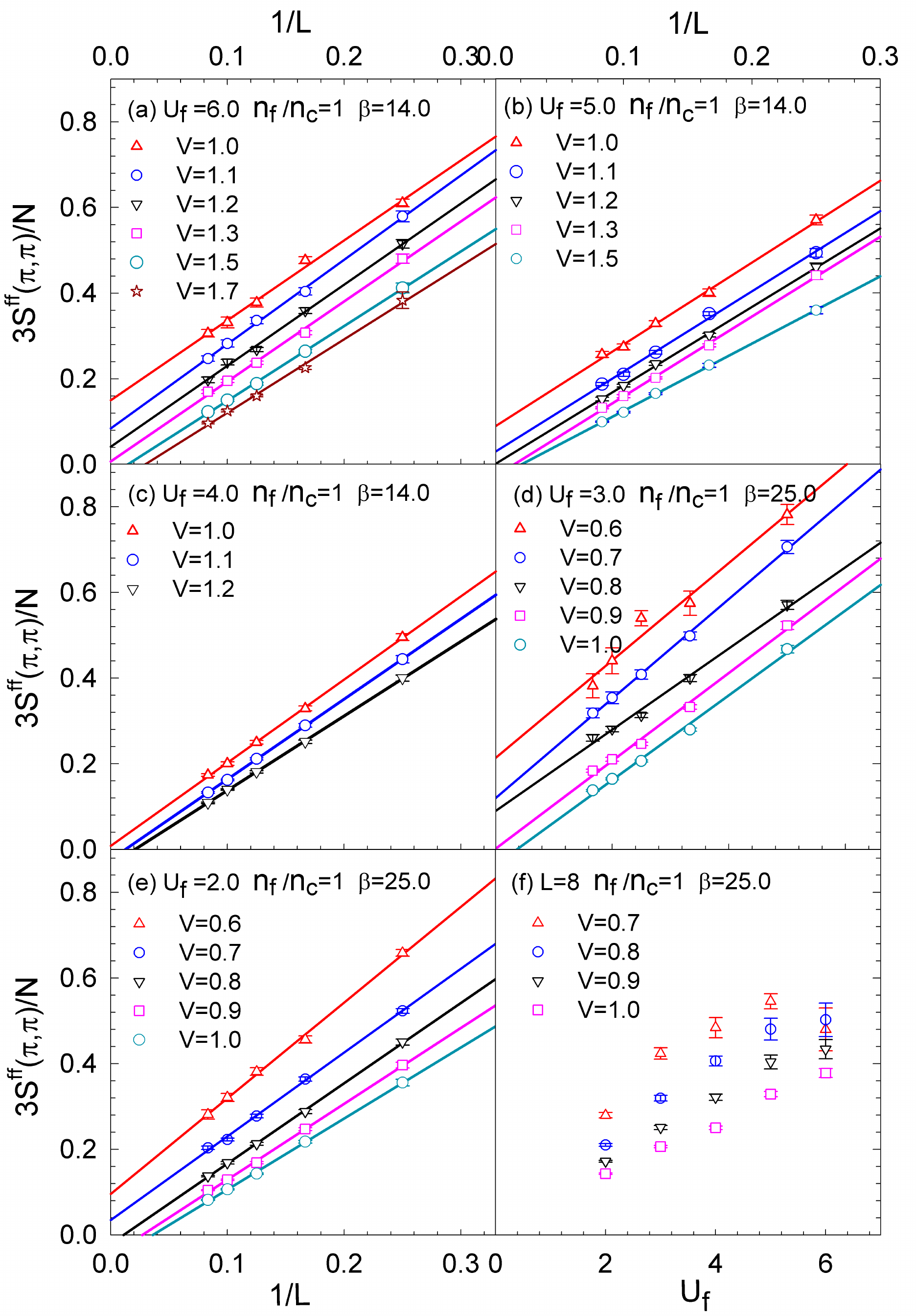}
\caption{($a$)--($e$):  Finite-size scaling of the structure factor
$S(\pi,\pi)$ when $q\equiv n_f/n_c=1$.
The data suggest values for the AF-singlet QCP
$V_{c} \simeq 1.3, 1.2, 1.0, 0.9, \text{and } 0.8$
(each estimate carries a rough error bar of 0.05),
for $U_f=6, 5, 4, 3, \text{and }2$, respectively.
These are consistent with the literature.
($f$) $S^\mathrm{ff}(\pi,\pi)$ vs $U_f$ for a fixed lattice
size $L=8$ and several values of $V$.
}
\label{Fig:Figure4}
\end{figure}

 \begin{figure}[tbp]
\includegraphics[scale=0.4]{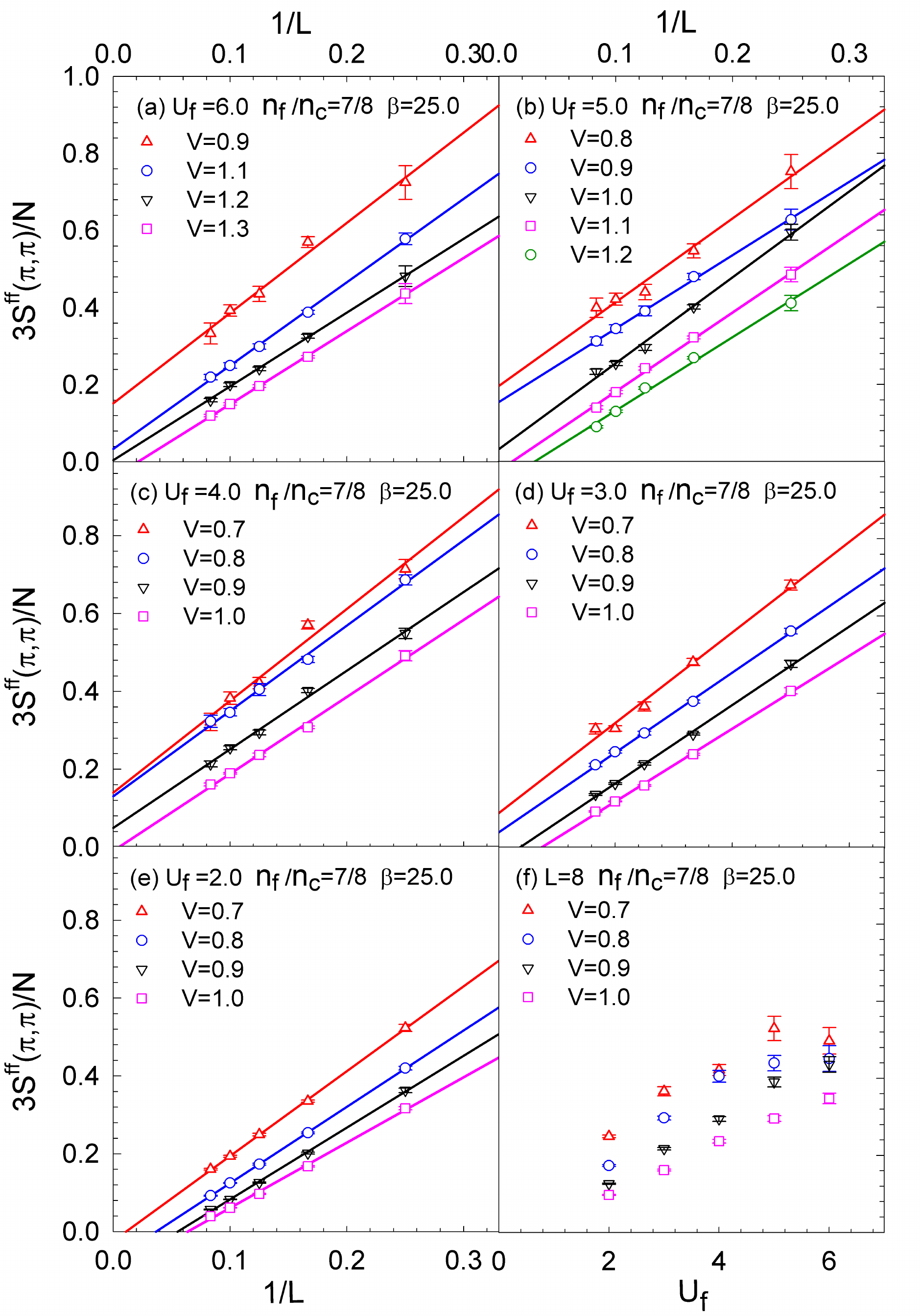}
\caption{($a$)--($e$)  Finite-size scaling of the
structure factor $S(\pi,\pi)$ when
local moments are removed from the lattice, as in
Fig.~\ref{Fig:Figure1}($b$).
The dilution fraction is $q\equiv n_f/n_c=7/8$.
The critical values for the AF-singlet QCP are somewhat
reduced by the lower density of magnetic
moments.  We estimate $V_{c} \simeq 1.2, 1.1,
0.9, 0.8, 0.6$
(each estimate carries a rough error bar of 0.05),
for $U_f=6, 5, 4, 3, 2$, respectively.
($f$) $S^\mathrm{ff}(\pi,\pi)$ vs $U_f$ for a fixed lattice
size $L=8$ and several values of $V$.
}
\label{Fig:Figure5}
\end{figure}

 \begin{figure}[tbp]
\includegraphics[scale=0.4]{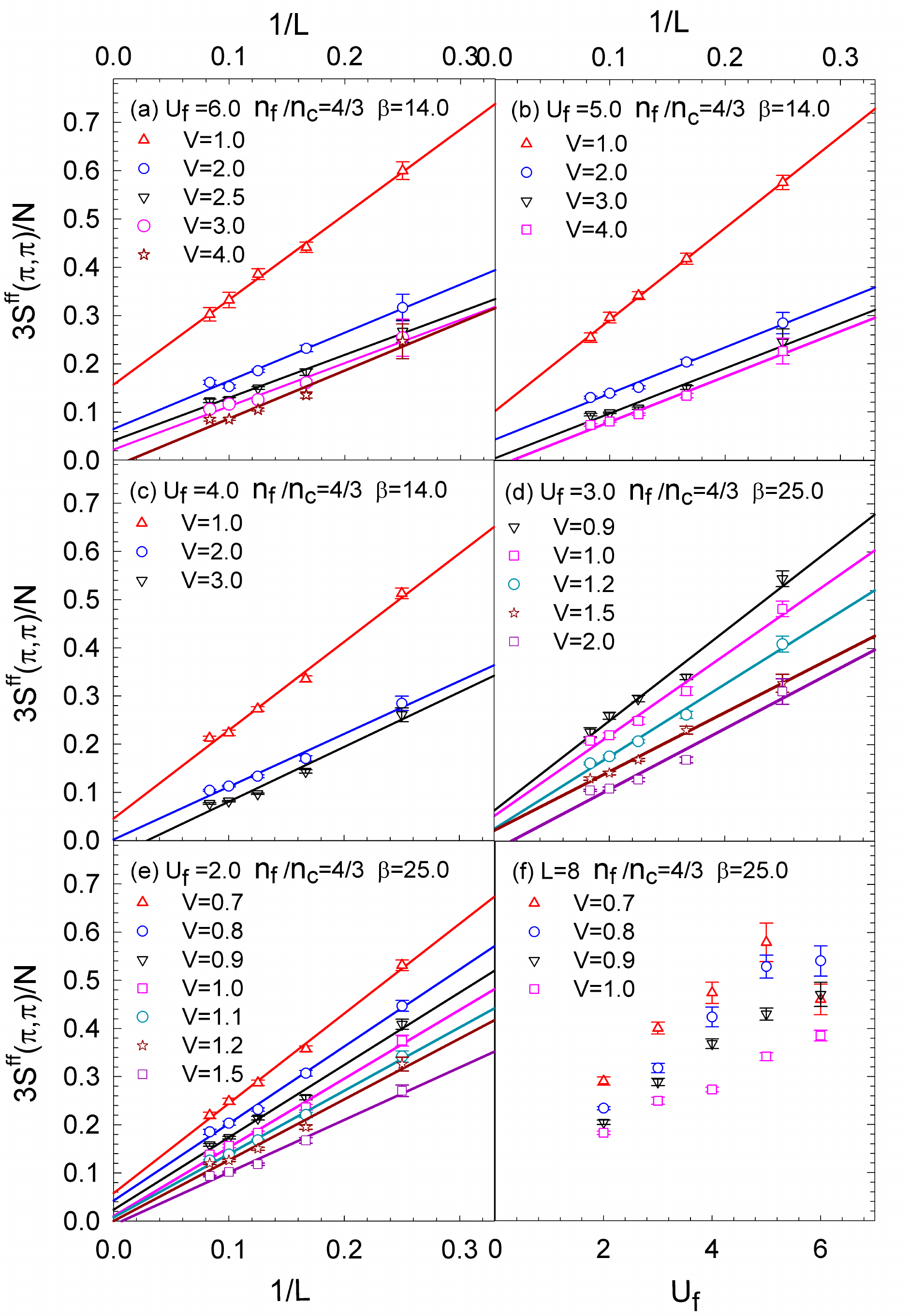}
\caption{($a$)--($e$):  Finite-size scaling of the structure factor
$S(\pi,\pi)$ when conduction orbitals are removed, as in
Fig.~\ref{Fig:Figure1}($c$).
The ratio of $f$ to $c$ orbitals is $q\equiv n_f/n_c=4/3$. Especially for larger $U_f=6, 5, 4$, ``exhaustion'' has
enhanced $V_{c}$ substantially: $V_{c}\simeq3.5, 3.0, 2.0, 1.5, 1.2$
(each estimate carries a rough error bar of 0.05),
for
$U_f=6.0, 5.0, 4.0, 3.0, 2.0$.  ($f$) $S^\mathrm{ff}(\pi,\pi)$ vs
$U_f$ for a fixed lattice size $L=8$ and several values of $V$.}
\label{Fig:Figure6}
\end{figure}


There are several additional features of Fig.~\ref{Fig:Figure3} which
are worth noting:  Most importantly, by comparing Figs. ~\ref{Fig:Figure3}(b) and ~\ref{Fig:Figure3}(e), we
see that dilution of conduction orbitals enhances the AF structure
factor, as it should.  Second, the AF structure factor decreases
as $U_f$ decreases, due to the magnitude of the local moments getting
smaller as a result of increasing charge fluctuations.  For instance,
$U_f=2$ requires larger $\beta$ to reach the ground state than $U_f=5$.
In fact, since the AF exchange in the Heisenberg limit is $J \sim
t^2/U_f$, a large $\beta$ is also required for $U_f \gg W = 8t$ (not
shown).

We probe the existence of long-range ordering by performing a
finite size scaling (FSS) analysis of  the structure factor. According
to spin-wave theory,\cite{huse88} the AF structure factor scales with
system size as
\begin{align}
\frac{1}{N} S^{\rm ff}(\pi,\pi) = \frac{1}{3} m^2 + \frac{a}{L} \, ,
\label{eq:fss}
\end{align}
with $m^2$ being the square of the AF order parameter, and $N=N_{f}$.
Figure ~\ref{Fig:Figure4} exhibits this FSS for different values of $U_f$
and $V$, for the undiluted PAM. It is interesting to notice that despite the NN hybridization, our results are similar to those of the
onsite case.\cite{vekic95,hu17} Our data suggest that the AF-singlet QCP
is located at $V_{c} \simeq$ 1.3, 1.2, 1.0, 0.9, and 0.8 (each estimate
carries a rough error bar of 0.05), for $U_f=$ 6, 5, 4, 3, and 2,
respectively. Also, similarly to the onsite case, $V_c$ for the
undiluted PAM is not too sensitive to the value of $U_f$; it only
changes by approximately 50\% over a range where $U_f$ is increased by a
factor of three.

We now turn our attention to the FSS analysis for the diluted
system, starting with the case $n_{f}<n_{c}$ (opposite to the exhaustion
limit); see, \textit{e.g.}, Fig. \ref{Fig:Figure1}(b).  Here we take
$q=n_f/n_c=7/8$, with the number of $c$ orbitals being $L^2$.
Following the preceding analyses,
Figs. \ref{Fig:Figure5}($a$)--\ref{Fig:Figure5}($e$) display the data for
different values of $U_{f}$ and $V$.  A comparison with
Fig.\,\ref{Fig:Figure4} reveals that the critical points $V_c(U_f)$ are
very close to those for the undiluted case.  As expected, the absence of
some local moments {\it reduces} the $V$ required to destroy AF order,
but the effect is around 10\%, \textit{i.e.}, of the order of
$1-n_f/n_c$.  This may be attributed to the longer-range character of
the effective RKKY interaction between the local moments.

On the other hand, the exhaustion scenario $n_f/n_c >1$ is dramatically different.
For instance, Fig.\ref{Fig:Figure6} presents the scaling analysis for
$n_f/n_c=4/3$, corresponding to one quarter of the conduction orbitals
removed.  It is quite evident that when the number of localized
electrons is larger than the conduction ones, the AF-singlet quantum
critical point is shifted to much larger values of $V$: long-range AF
ordering is stabilized for $V \lesssim 2$ when $U_f=4$, and for $V
\lesssim 3$ when $U_f=5$.  This should be contrasted with the small
changes [relative to the undiluted $V_c(U_f)$] that occur when $n_f/n_c
< 1$ or which accompany altering $U_f$ at $n_c/n_f=1$.

\begin{figure}[t!]
\includegraphics[scale=0.5]{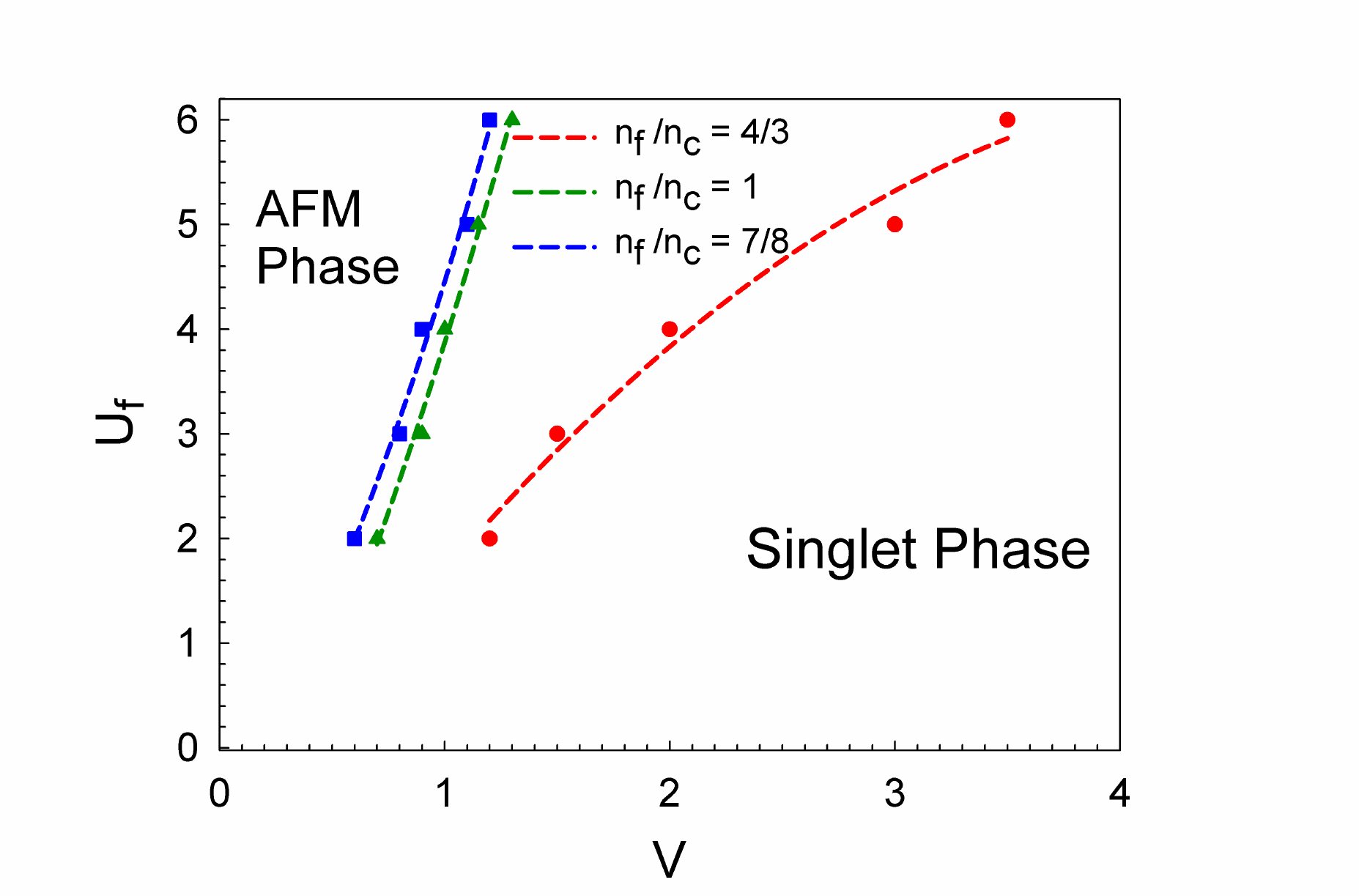}
\caption{
The ground-state phase diagram.
Green triangles mark the AF-singlet transition boundary at
$n_f/n_c=1$, the conventional (i.e.~undiluted) periodic Anderson model.
The blue squares indicate the boundary for $n_f/n_c=7/8$, with dilution
of {\it local} electrons. Very little change is noted.
Red circles indicate the boundary for $n_f/n_c=4/3$, dilution
of {\it conduction} electrons, when exhaustion
is present. In this case, the stability of AF is dramatically
increased.
}
\label{Fig:Figure7}
\end{figure}

Figures ~\ref{Fig:Figure4}$(f)$, \ref{Fig:Figure5}$(f)$, and
\ref{Fig:Figure6}$(f)$ share the feature that the AF structure factor
$S^{\rm ff}(\pi,\pi)$ grows monotonically with $U_f$.  We expect that at fixed
inverse temperature $\beta$, these curves will eventually turn over and
decrease, owing to the $1/U_f$ behavior of the exchange constant at
strong coupling.  In the single-band Hubbard model, maximal AF
correlations occur at $U_f \sim 8$ for $\beta \sim
12$.\cite{scalettar91} In summary, from these preceding results, we
observe that the critical hybridization for an AF-singlet transition
grows dramatically when the conduction-electron count is smaller than
the localized ones, while $V_{c}(U_{f})$ is only weakly changed in the opposite
situation.  All results for the QCP are summarized in the phase diagram
of Fig. \ref{Fig:Figure7}.

\begin{figure}[tbp]
\includegraphics[scale=0.4]{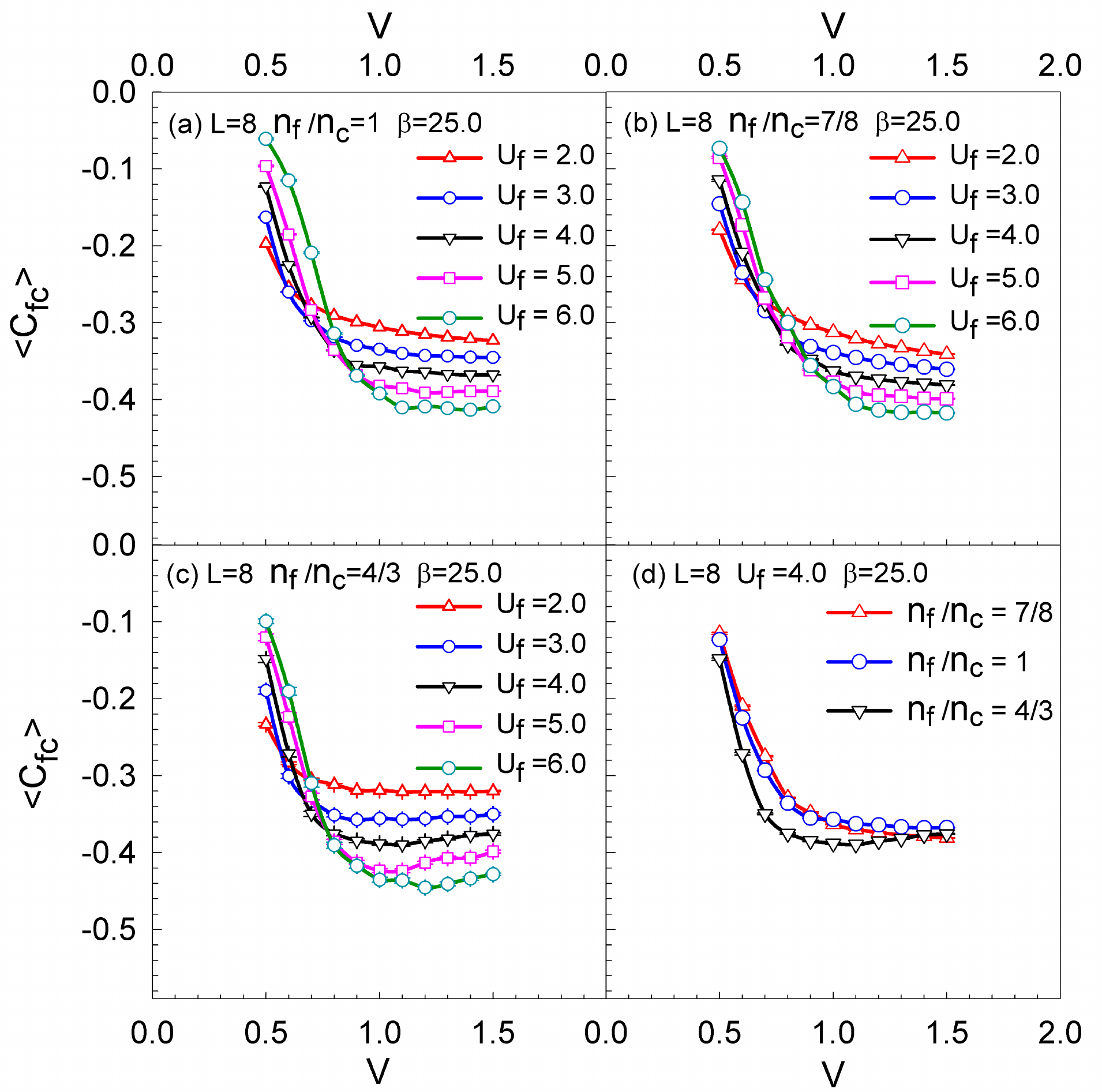}
\caption{
($a$)--($c$) The behavior of the singlet correlator as a function of $V$ for different $U_f$ and the three ratios $n_f/n_c=1,
7/8, 4/3$.  The lattice size $L=8$ and inverse temperature $\beta=25$.
There is a general tendency for singlet formation to occur at $V \sim
0.5$-$1.0$ for all three filling ratios, as emphasized in ($d$).
}
\label{Fig:Figure8}
\end{figure}

We have also investigated the singlet formation by calculating the local
singlet correlator, given by Eq. \eqref{eq:Cfd}. Figure \ref{Fig:Figure8} shows
$C_{fc}$ as a function of $V$, that is, as the AF-singlet transition is
traversed, for different values of $U_f$.  Figures \ref{Fig:Figure8}$(a)$--\ref{Fig:Figure8}$(c)$ display
results for several values of $U_f$, and for the filling ratios
$n_f/n_c=$1, 7/8, and 4/3, respectively.  For these cases, $C_{fc}$
increases in magnitude, from small values to $|C_{fc}| \sim 0.4$, as $V$
changes from $V\sim 0.5$ to $V\sim 1.0$.  The curves for the three
filling ratios also all exhibit a crossing pattern:  At weak
hybridization $V$, $C_{fc}$ is largest in magnitude at weak coupling
$U_f=2$.  However, as $V$ increases, $C_{fc}$ becomes largest in
magnitude at $U_f=6$;  we interpret this as occurring because large $U_f$
yields the most well-formed moments on the $f$ sites.

We have seen that $V_c$ for the destruction of AF order increases
dramatically for the exhaustion value, $n_f/n_c=4/3$, when $n_f$ exceeds
$n_c$.  It is intriguing that, in Fig.\,\ref{Fig:Figure8} there is not
as great a reflection of this in the values of $V$ at which singlet
correlators develop.  That is, $|C_{fc}|$ grows from small values to
$|C_{\rm fc}| \sim 0.4$ in the same range $0.5 \lesssim V \lesssim 1.0$ for
 $n_f/n_c=4/3$ as for $n_f/n_c=7/8$ and 1.  It appears, therefore, that
the interval from $V \sim 1$ to $V_c \sim 3$ (for $U_f = 5,6$) is
characterized by relatively large values of the singlet correlator, even
though AF order remains.  Whether it corresponds to a partially screened
region with coexistence between singlet and AF is a challenge to
resolve conclusively with the DQMC methodology used here.
While not evident in the onset of $C_{fc}$,
the enhancement of $V_c$ by exhaustion appears to be
reflected in the nonmonotonic evolution of $C_{fc}$, which is unique to
the $n_f/n_c=4/3$ case; see, \textit{e.g.},
Fig. \ref{Fig:Figure8}($c$). The similarity of evolution of the onset
for the three filling ratios is emphasized by replotting the data of
Figs. \ref{Fig:Figure8}($a$)--\ref{Fig:Figure8}($c$) for all three ratios (at a single value $U_f=4$) in
Fig. \ref{Fig:Figure8}(d).

 \begin{figure}[tbp]
\includegraphics[scale=0.4]{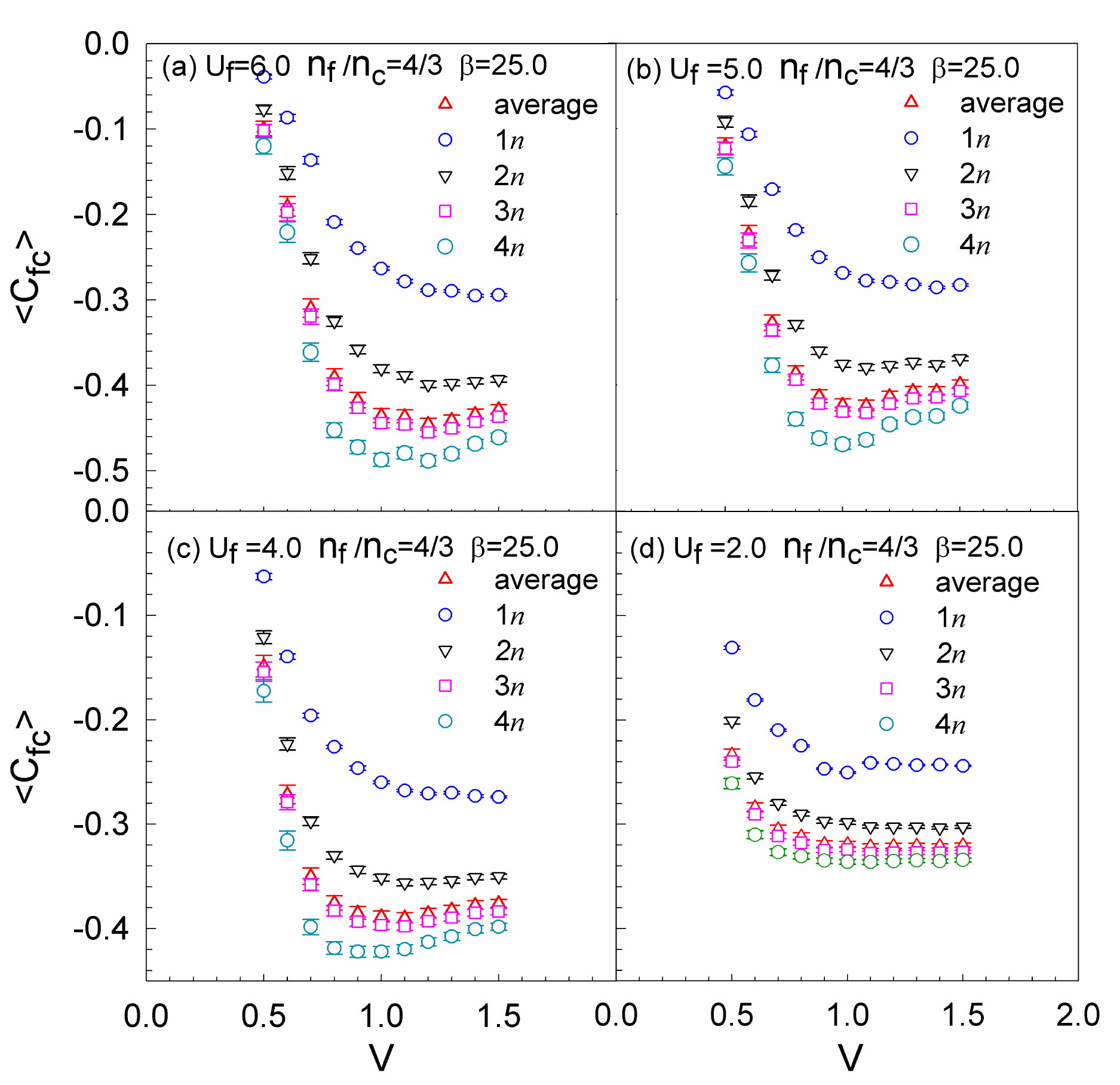}
\caption{
The singlet correlators for different numbers
of conduction electron neighbors are shown as a function of $V$ for
the filling ratio $n_f/n_c=4/3$ which realizes exhaustion. $1n$, $2n$, $3n$, $4n$ correspond to $f$ orbitals connected to 1, 2, 3, and 4
conduction orbitals by $V$.
}
\label{Fig:Figure9}
\end{figure}
 \begin{figure}[tbp]
\includegraphics[scale=0.4]{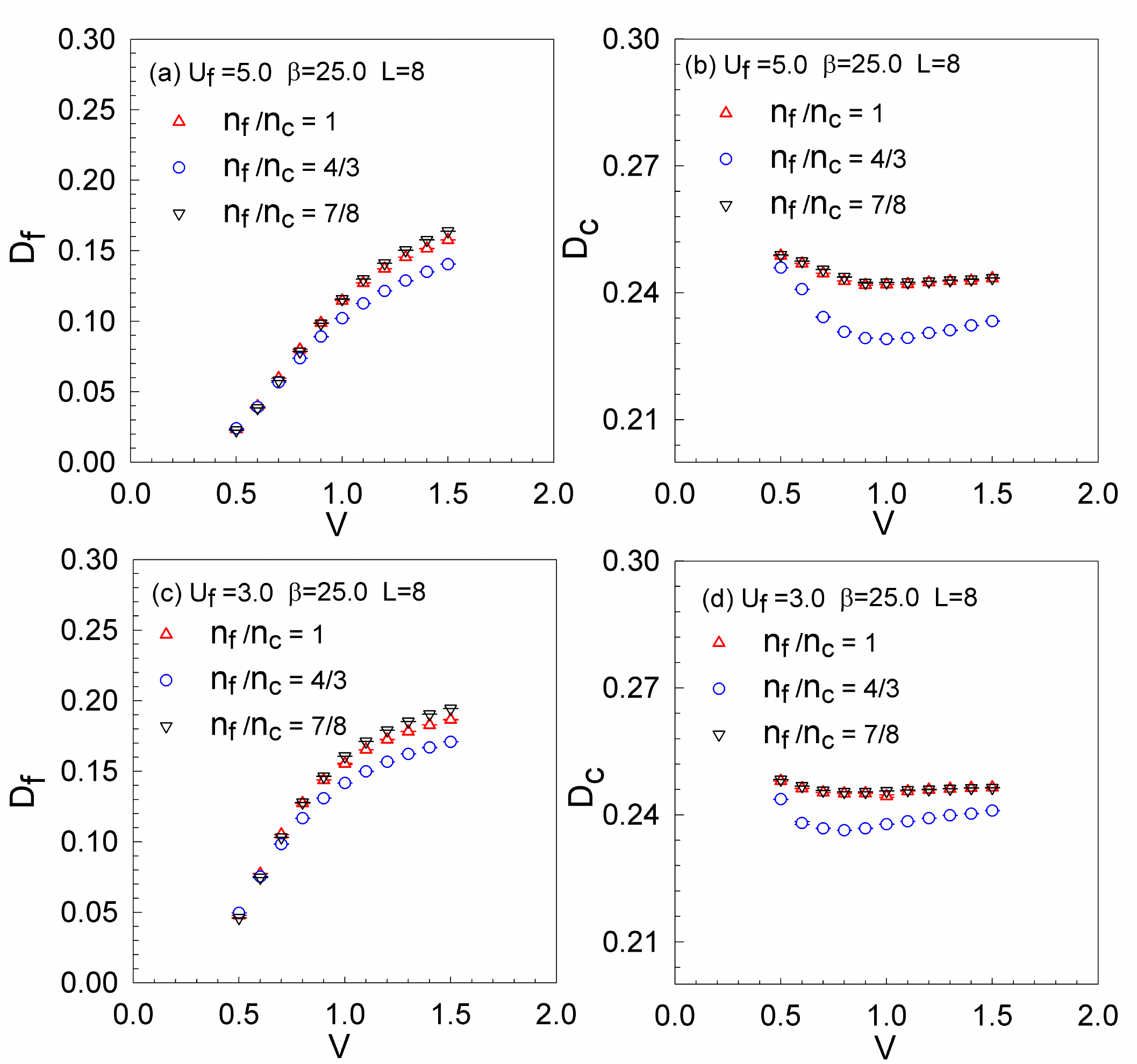}
\caption{
Double occupancy $D_f(D_c)$ on the $f(c)$ sites vs
hybridization $V$.  Larger $V$ results in an increase in $D_f$.
Conduction-electron dilution ($n_f/n_c=4/3$) is correlated with smaller
$D_f$:  there are fewer conduction electrons to hop onto the $f$
orbitals.  $D_c$ is roughly at the noninteracting value $1/4$ for all
situations.
}
\label{Fig:Figure10}
\end{figure}

The data shown in Fig. \ref{Fig:Figure8} result from averaging $C_{fc}$
over all $f$ orbitals.  However, as discussed in relation to
Eq. \eqref{eq:Pm}, diluting conduction electrons leads to nonequivalent
$f$ orbitals, depending on the number of active connected $c$ sites.
Data for $C_{fc}$ can therefore be decomposed according to whether this
number is 1, 2, 3, or 4 (the largest value for a square lattice with only
near-neighbor $f$-$c$ hopping).  These are, respectively, denoted by $1n, 2n, 3n, 4n$
in Fig. \ref{Fig:Figure9}. We show this only for the
case of exhaustion, $n_f/n_c=4/3$.  As it might have been intuitively
expected, $C_{fc}$ is largest in magnitude for $4n$, and smallest for
$1n$.  It appears that $C_{fc}$ also begins to grow in magnitude at
smaller $V$ for $4n$ than for $1n$.

Fig.\ref{Fig:Figure10} shows the behavior of the double occupancy,
$D_{\alpha}=\langle n^{\alpha}_{\bf i\uparrow} n^{\alpha}_{\bf i\downarrow} \rangle$, $\alpha=c$ or $f$, with $V$ for $U_f=3$ and $U_f=5$;
the left panels show $D_f$, while and the right panels $D_c$.
The onsite repulsion $U_f$ makes $D_f$ small, especially at small $V$ where quantum fluctuations are suppressed.
By contrast, $D_c$ changes very little over the range of $V$ examined,
taking on values close to the uncorrelated limit, $D_c \sim \langle
n^d_{\bf i\uparrow}  \rangle \langle n^d_{\bf i\downarrow} \rangle \sim
1/4$.

\section{Conclusions} \label{Sec:Conclusions}

One of the quantitative conclusions of past QMC studies of the PAM (in
its different variants) is that the position of the QCP's,
which signal the AF-singlet transition in the ground state,
is rather weakly dependent on the parameters of the model, especially on
the value of the on-site repulsion $U_f$.
This was noted in Refs.\,\onlinecite{vekic95,hu17}, where the phase
boundaries were found to be rather vertical in the $U_f-V$ plane, in
contrast with mean field theory predictions of much larger $dV_c/dU_f$.
Neither was $V_c$ found to vary much in going from 2D to
3D,\cite{huscroft99} or with changes in the momentum dependence of the
$f$-$c$ hybridization.\cite{huscroft99,held00}  In the former case, the
noninteracting density of states is divergent at half-filling for 2D,
and finite for 3D.  In the latter case, the on-site and intersite forms for
$V_{\bf k}$ give rise to very different band structures:  insulating for
$V$ independent of ${\bf k}$, and metallic for nearest neighbor hybridization.  Despite
these seemingly important differences, $V_c$ was found to be not only
rather immune to changes in $U_f$, but also insensitive to the
underlying band structure.

By contrast, one of our key results here, then, is that $V_c$ can vary dramatically
with the ratio $n_c/n_f$.
This effect is summarized in the phase diagram in Fig.~6.  The physics
of exhaustion is starkly evident.
While dilution of {\it local} electrons barely shifts the phase boundary (singlet formation occurs
slightly earlier), dilution of {\it conduction} electrons delays singlet formation to values of $V$ as much as a factor of three greater than in the balanced case, the conventional PAM.
This result has the potential to lend qualitative insight into heavy
fermion materials since their doping can proceed both by the elimination
of moments, {\it e.g.}~Ce$_{1-x}$La$_x$CoIn$_5$, and also by changes to
the conduction electrons {\it e.g.}~CeCo$_{1-x}$Cd$_x$In$_5$.
Theoretical descriptions of the latter situation have focused on the
impurity-induced changes to the hybridization $V$ rather than changes to
$n_c$.

A second key, and rather unexpected, conclusion is that in the interval $1 \lesssim V \lesssim 3$ in which exhaustion induces AF order (for $U_f= 5$) the singlet correlator is large.
In the undiluted case the same $V_c$ pinpoints where the AF structure factor becomes small and the singlet correlator becomes large.
In the presence of exhaustion, however, these two events no longer share a common $V_c$, and there is an extended region where both $|C_{fc}|$ is large {\it and} AF order is still present.

The large increase in $V_c$ found for $n_f/n_c>1$ reflects a significant
increase in the stability of the AF phase,
the mechanism of which can be attributed to
the increased difficulty of forming singlets when the number of conduction
electrons available for screening is reduced.
Although we have not
evaluated $T_K$ and $T_{\rm coh}$ here (it is possible to do so with QMC
in impurity limit using, e.g., the Hirsch-Fye method,\cite{hirsch86} but
much harder for the lattice) it is reasonable to suppose that these
temperatures will be reduced to reflect the decreased tendency towards
forming singlets.  We have also emphasized that the ground state
structure factor $S^{\rm ff}(\pi,\pi)$ is larger for $n_f/n_c=4/3$ than for
$n_f/n_c=1$, consistent with the general trend towards stronger
AF with (a moderate degree of) exhaustion.

The focus of this paper has been on the competition of AF order and singlet formation (for both the cases $n_f/n_c>1$ and
$n_f/n_c<1$). The physics of exhaustion appears as a large increase in
AF stability. It is worth noting that the PAM has also been extensively
studied as a model of ferromagnetism (FM).\cite{meyer00b,batista03}
The Nagaoka theorem notwithstanding, FM appears to be very difficult to
be achieved in single band systems such as those described by the Hubbard
Hamiltonian. Exploration of $\mathbf{q}=0$ order would be an interesting avenue
to pursue, {\it e.g.}~in a more extreme limit $n_f/n_c \gg 1$ than that
considered here.

\noindent
\underline{\it Acknowledgement} ---
L.Z. and T.M. were supported by NSFC (Grant No. 11774033) and Beijing Natural Science Foundation (Grant No. 1192011).
The numerical simulations in this work were performed on the HSCC of
Beijing Normal University and Tianhe in Beijing Computational Science Research Center.
N.C.C and R.R.dS. acknowledge support by the Brazilian agencies CAPES, CNPq, and FAPERJ.
The work of R.T.S was supported by Grant No. DOE-DE-SC0014671.

\bibliography{reference}

\begin{thebibliography}{75}%
\makeatletter
\providecommand \@ifxundefined [1]{%
 \@ifx{#1\undefined}
}%
\providecommand \@ifnum [1]{%
 \ifnum #1\expandafter \@firstoftwo
 \else \expandafter \@secondoftwo
 \fi
}%
\providecommand \@ifx [1]{%
 \ifx #1\expandafter \@firstoftwo
 \else \expandafter \@secondoftwo
 \fi
}%
\providecommand \natexlab [1]{#1}%
\providecommand \enquote  [1]{``#1''}%
\providecommand \bibnamefont  [1]{#1}%
\providecommand \bibfnamefont [1]{#1}%
\providecommand \citenamefont [1]{#1}%
\providecommand \href@noop [0]{\@secondoftwo}%
\providecommand \href [0]{\begingroup \@sanitize@url \@href}%
\providecommand \@href[1]{\@@startlink{#1}\@@href}%
\providecommand \@@href[1]{\endgroup#1\@@endlink}%
\providecommand \@sanitize@url [0]{\catcode `\\12\catcode `\$12\catcode
  `\&12\catcode `\#12\catcode `\^12\catcode `\_12\catcode `\%12\relax}%
\providecommand \@@startlink[1]{}%
\providecommand \@@endlink[0]{}%
\providecommand \url  [0]{\begingroup\@sanitize@url \@url }%
\providecommand \@url [1]{\endgroup\@href {#1}{\urlprefix }}%
\providecommand \urlprefix  [0]{URL }%
\providecommand \Eprint [0]{\href }%
\providecommand \doibase [0]{http://dx.doi.org/}%
\providecommand \selectlanguage [0]{\@gobble}%
\providecommand \bibinfo  [0]{\@secondoftwo}%
\providecommand \bibfield  [0]{\@secondoftwo}%
\providecommand \translation [1]{[#1]}%
\providecommand \BibitemOpen [0]{}%
\providecommand \bibitemStop [0]{}%
\providecommand \bibitemNoStop [0]{.\EOS\space}%
\providecommand \EOS [0]{\spacefactor3000\relax}%
\providecommand \BibitemShut  [1]{\csname bibitem#1\endcsname}%
\let\auto@bib@innerbib\@empty
\bibitem [{\citenamefont {Anderson}(1961)}]{anderson61}%
  \BibitemOpen
  \bibfield  {author} {\bibinfo {author} {\bibfnamefont {P.~W.}\ \bibnamefont
  {Anderson}},\ }\href {\doibase 10.1103/PhysRev.124.41} {\bibfield  {journal}
  {\bibinfo  {journal} {Phys. Rev.}\ }\textbf {\bibinfo {volume} {124}},\
  \bibinfo {pages} {41} (\bibinfo {year} {1961})}\BibitemShut {NoStop}%
\bibitem [{\citenamefont {Stewart}(1984)}]{stewart84}%
  \BibitemOpen
  \bibfield  {author} {\bibinfo {author} {\bibfnamefont {G.~R.}\ \bibnamefont
  {Stewart}},\ }\href {\doibase 10.1103/RevModPhys.56.755} {\bibfield
  {journal} {\bibinfo  {journal} {Rev. Mod. Phys.}\ }\textbf {\bibinfo {volume}
  {56}},\ \bibinfo {pages} {755} (\bibinfo {year} {1984})}\BibitemShut
  {NoStop}%
\bibitem [{\citenamefont {Nozieres}(1985)}]{nozieres85}%
  \BibitemOpen
  \bibfield  {author} {\bibinfo {author} {\bibfnamefont {P.}~\bibnamefont
  {Nozieres}},\ }\href {https://doi.org/10.1051/anphys:0198500100101900}
  {\bibfield  {journal} {\bibinfo  {journal} {Ann. De Phys.}\ }\textbf
  {\bibinfo {volume} {10}},\ \bibinfo {pages} {19} (\bibinfo {year}
  {1985})}\BibitemShut {NoStop}%
\bibitem [{\citenamefont {Lee}\ \emph {et~al.}(1986)\citenamefont {Lee},
  \citenamefont {Rice}, \citenamefont {Serene}, \citenamefont {Sham},\ and\
  \citenamefont {Wilkins}}]{lee86}%
  \BibitemOpen
  \bibfield  {author} {\bibinfo {author} {\bibfnamefont {P.~A.}\ \bibnamefont
  {Lee}}, \bibinfo {author} {\bibfnamefont {T.~M.}\ \bibnamefont {Rice}},
  \bibinfo {author} {\bibfnamefont {J.~W.}\ \bibnamefont {Serene}}, \bibinfo
  {author} {\bibfnamefont {L.~J.}\ \bibnamefont {Sham}}, \ and\ \bibinfo
  {author} {\bibfnamefont {J.~W.}\ \bibnamefont {Wilkins}},\ }\href
  {https://inis.iaea.org/search/search.aspx?orig_q=RN:17058516} {\bibfield
  {journal} {\bibinfo  {journal} {Comments on Condensed Matter Physics}\
  }\textbf {\bibinfo {volume} {12}},\ \bibinfo {pages} {99} (\bibinfo {year}
  {1986})}\BibitemShut {NoStop}%
\bibitem [{\citenamefont {Andrei}\ \emph {et~al.}(1983)\citenamefont {Andrei},
  \citenamefont {Furuya},\ and\ \citenamefont {Lowenstein}}]{andrei83}%
  \BibitemOpen
  \bibfield  {author} {\bibinfo {author} {\bibfnamefont {N.}~\bibnamefont
  {Andrei}}, \bibinfo {author} {\bibfnamefont {K.}~\bibnamefont {Furuya}}, \
  and\ \bibinfo {author} {\bibfnamefont {J.~H.}\ \bibnamefont {Lowenstein}},\
  }\href {\doibase 10.1103/RevModPhys.55.331} {\bibfield  {journal} {\bibinfo
  {journal} {Rev. Mod. Phys.}\ }\textbf {\bibinfo {volume} {55}},\ \bibinfo
  {pages} {331} (\bibinfo {year} {1983})}\BibitemShut {NoStop}%
\bibitem [{\citenamefont {Schlottmann}(1987)}]{schlottmann87}%
  \BibitemOpen
  \bibfield  {author} {\bibinfo {author} {\bibfnamefont {P.}~\bibnamefont
  {Schlottmann}},\ }\href {\doibase 10.1103/PhysRevB.36.5177} {\bibfield
  {journal} {\bibinfo  {journal} {Phys. Rev. B}\ }\textbf {\bibinfo {volume}
  {36}},\ \bibinfo {pages} {5177} (\bibinfo {year} {1987})}\BibitemShut
  {NoStop}%
\bibitem [{\citenamefont {Ruderman}\ and\ \citenamefont
  {Kittel}(1954)}]{ruderman54}%
  \BibitemOpen
  \bibfield  {author} {\bibinfo {author} {\bibfnamefont {M.~A.}\ \bibnamefont
  {Ruderman}}\ and\ \bibinfo {author} {\bibfnamefont {C.}~\bibnamefont
  {Kittel}},\ }\href {\doibase 10.1103/PhysRev.96.99} {\bibfield  {journal}
  {\bibinfo  {journal} {Phys. Rev.}\ }\textbf {\bibinfo {volume} {96}},\
  \bibinfo {pages} {99} (\bibinfo {year} {1954})}\BibitemShut {NoStop}%
\bibitem [{\citenamefont {Kasuya}(1956)}]{kasuya56}%
  \BibitemOpen
  \bibfield  {author} {\bibinfo {author} {\bibfnamefont {T.}~\bibnamefont
  {Kasuya}},\ }\href {\doibase 10.1143/PTP.16.45} {\bibfield  {journal}
  {\bibinfo  {journal} {Progress of Theoretical Physics}\ }\textbf {\bibinfo
  {volume} {16}},\ \bibinfo {pages} {45} (\bibinfo {year} {1956})}\BibitemShut
  {NoStop}%
\bibitem [{\citenamefont {Yosida}(1957)}]{yosida57}%
  \BibitemOpen
  \bibfield  {author} {\bibinfo {author} {\bibfnamefont {K.}~\bibnamefont
  {Yosida}},\ }\href {\doibase 10.1103/PhysRev.106.893} {\bibfield  {journal}
  {\bibinfo  {journal} {Phys. Rev.}\ }\textbf {\bibinfo {volume} {106}},\
  \bibinfo {pages} {893} (\bibinfo {year} {1957})}\BibitemShut {NoStop}%
\bibitem [{\citenamefont {Xavier}\ \emph {et~al.}(2004)\citenamefont {Xavier},
  \citenamefont {Miranda},\ and\ \citenamefont {Dagotto}}]{Xavier04}%
  \BibitemOpen
  \bibfield  {author} {\bibinfo {author} {\bibfnamefont {J.~C.}\ \bibnamefont
  {Xavier}}, \bibinfo {author} {\bibfnamefont {E.}~\bibnamefont {Miranda}}, \
  and\ \bibinfo {author} {\bibfnamefont {E.}~\bibnamefont {Dagotto}},\ }\href
  {\doibase 10.1103/PhysRevB.70.172415} {\bibfield  {journal} {\bibinfo
  {journal} {Phys. Rev. B}\ }\textbf {\bibinfo {volume} {70}},\ \bibinfo
  {pages} {172415} (\bibinfo {year} {2004})}\BibitemShut {NoStop}%
\bibitem [{\citenamefont {Peters}\ and\ \citenamefont
  {Kawakami}(2015)}]{Peters15}%
  \BibitemOpen
  \bibfield  {author} {\bibinfo {author} {\bibfnamefont {R.}~\bibnamefont
  {Peters}}\ and\ \bibinfo {author} {\bibfnamefont {N.}~\bibnamefont
  {Kawakami}},\ }\href {\doibase 10.1103/PhysRevB.92.075103} {\bibfield
  {journal} {\bibinfo  {journal} {Phys. Rev. B}\ }\textbf {\bibinfo {volume}
  {92}},\ \bibinfo {pages} {075103} (\bibinfo {year} {2015})}\BibitemShut
  {NoStop}%
\bibitem [{\citenamefont {Costa}\ \emph {et~al.}(2017)\citenamefont {Costa},
  \citenamefont {Lima},\ and\ \citenamefont {\surname{dos Santos}}}]{costa17}%
  \BibitemOpen
  \bibfield  {author} {\bibinfo {author} {\bibfnamefont {N.~C.}\ \bibnamefont
  {Costa}}, \bibinfo {author} {\bibfnamefont {J.~P.}\ \bibnamefont {Lima}}, \
  and\ \bibinfo {author} {\bibfnamefont {R.~R.}\ \bibnamefont {\surname{dos
  Santos}}},\ }\href {\doibase 10.1016/j.jmmm.2016.09.061} {\bibfield
  {journal} {\bibinfo  {journal} {J. of Mag. and Mag. Mat.}\ }\textbf {\bibinfo
  {volume} {423}},\ \bibinfo {pages} {74} (\bibinfo {year} {2017})}\BibitemShut
  {NoStop}%
\bibitem [{\citenamefont {Igoshev}\ \emph {et~al.}(2017)\citenamefont
  {Igoshev}, \citenamefont {Timirgazin}, \citenamefont {Arzhnikov},
  \citenamefont {Antipin},\ and\ \citenamefont {Irkhin}}]{Igoshev17}%
  \BibitemOpen
  \bibfield  {author} {\bibinfo {author} {\bibfnamefont {P.}~\bibnamefont
  {Igoshev}}, \bibinfo {author} {\bibfnamefont {M.}~\bibnamefont {Timirgazin}},
  \bibinfo {author} {\bibfnamefont {A.}~\bibnamefont {Arzhnikov}}, \bibinfo
  {author} {\bibfnamefont {T.}~\bibnamefont {Antipin}}, \ and\ \bibinfo
  {author} {\bibfnamefont {V.}~\bibnamefont {Irkhin}},\ }\href {\doibase
  https://doi.org/10.1016/j.jmmm.2016.12.064} {\bibfield  {journal} {\bibinfo
  {journal} {Journal of Magnetism and Magnetic Materials}\ }\textbf {\bibinfo
  {volume} {440}},\ \bibinfo {pages} {66 } (\bibinfo {year}
  {2017})}\BibitemShut {NoStop}%
\bibitem [{\citenamefont {Zhong}\ \emph {et~al.}(2019)\citenamefont {Zhong},
  \citenamefont {Yang}, \citenamefont {Zhao},\ and\ \citenamefont
  {Luo}}]{Zhong19}%
  \BibitemOpen
  \bibfield  {author} {\bibinfo {author} {\bibfnamefont {Y.}~\bibnamefont
  {Zhong}}, \bibinfo {author} {\bibfnamefont {W.-W.}\ \bibnamefont {Yang}},
  \bibinfo {author} {\bibfnamefont {J.-Z.}\ \bibnamefont {Zhao}}, \ and\
  \bibinfo {author} {\bibfnamefont {H.-G.}\ \bibnamefont {Luo}},\ }\href
  {https://arxiv.org/abs/1903.05295v2} {\bibfield  {journal} {\bibinfo
  {journal} {arXiv:1903.05295}\ } (\bibinfo {year} {2019})}\BibitemShut
  {NoStop}%
\bibitem [{\citenamefont {Vidhyadhiraja}\ and\ \citenamefont
  {Logan}(2004)}]{vidhyadhiraja04}%
  \BibitemOpen
  \bibfield  {author} {\bibinfo {author} {\bibfnamefont {N.~S.}\ \bibnamefont
  {Vidhyadhiraja}}\ and\ \bibinfo {author} {\bibfnamefont {D.~E.}\ \bibnamefont
  {Logan}},\ }\href {\doibase 10.1140/epjb/e2004-00197-6} {\bibfield  {journal}
  {\bibinfo  {journal} {Eur. Phys. J. B}\ }\textbf {\bibinfo {volume} {39}},\
  \bibinfo {pages} {313} (\bibinfo {year} {2004})}\BibitemShut {NoStop}%
\bibitem [{\citenamefont {Hewson}(1993)}]{hewson93}%
  \BibitemOpen
  \bibfield  {author} {\bibinfo {author} {\bibfnamefont {A.~C.}\ \bibnamefont
  {Hewson}},\ }\href {\doibase 10.1017/CBO9780511470752} {\emph {\bibinfo
  {title} {The Kondo Problem to Heavy Fermions}}},\ Cambridge Studies in
  Magnetism\ (\bibinfo  {publisher} {Cambridge University Press},\ \bibinfo
  {year} {1993})\BibitemShut {NoStop}%
\bibitem [{\citenamefont {Tsunetsugu}\ \emph {et~al.}(1997)\citenamefont
  {Tsunetsugu}, \citenamefont {Sigrist},\ and\ \citenamefont
  {Ueda}}]{tsunetsugu97}%
  \BibitemOpen
  \bibfield  {author} {\bibinfo {author} {\bibfnamefont {H.}~\bibnamefont
  {Tsunetsugu}}, \bibinfo {author} {\bibfnamefont {M.}~\bibnamefont {Sigrist}},
  \ and\ \bibinfo {author} {\bibfnamefont {K.}~\bibnamefont {Ueda}},\ }\href
  {\doibase 10.1103/RevModPhys.69.809} {\bibfield  {journal} {\bibinfo
  {journal} {Rev. Mod. Phys.}\ }\textbf {\bibinfo {volume} {69}},\ \bibinfo
  {pages} {809} (\bibinfo {year} {1997})}\BibitemShut {NoStop}%
\bibitem [{\citenamefont {Rice}\ and\ \citenamefont {Ueda}(1985)}]{rice85}%
  \BibitemOpen
  \bibfield  {author} {\bibinfo {author} {\bibfnamefont {T.~M.}\ \bibnamefont
  {Rice}}\ and\ \bibinfo {author} {\bibfnamefont {K.}~\bibnamefont {Ueda}},\
  }\href {\doibase 10.1103/PhysRevLett.55.995} {\bibfield  {journal} {\bibinfo
  {journal} {Phys. Rev. Lett.}\ }\textbf {\bibinfo {volume} {55}},\ \bibinfo
  {pages} {995} (\bibinfo {year} {1985})}\BibitemShut {NoStop}%
\bibitem [{\citenamefont {Rice}\ and\ \citenamefont {Ueda}(1986)}]{rice86}%
  \BibitemOpen
  \bibfield  {author} {\bibinfo {author} {\bibfnamefont {T.~M.}\ \bibnamefont
  {Rice}}\ and\ \bibinfo {author} {\bibfnamefont {K.}~\bibnamefont {Ueda}},\
  }\href {\doibase 10.1103/PhysRevB.34.6420} {\bibfield  {journal} {\bibinfo
  {journal} {Phys. Rev. B}\ }\textbf {\bibinfo {volume} {34}},\ \bibinfo
  {pages} {6420} (\bibinfo {year} {1986})}\BibitemShut {NoStop}%
\bibitem [{\citenamefont {Veki\ifmmode~\acute{c}\else \'{c}\fi{}}\ \emph
  {et~al.}(1995)\citenamefont {Veki\ifmmode~\acute{c}\else \'{c}\fi{}},
  \citenamefont {Cannon}, \citenamefont {Scalapino}, \citenamefont
  {Scalettar},\ and\ \citenamefont {Sugar}}]{vekic95}%
  \BibitemOpen
  \bibfield  {author} {\bibinfo {author} {\bibfnamefont {M.}~\bibnamefont
  {Veki\ifmmode~\acute{c}\else \'{c}\fi{}}}, \bibinfo {author} {\bibfnamefont
  {J.~W.}\ \bibnamefont {Cannon}}, \bibinfo {author} {\bibfnamefont {D.~J.}\
  \bibnamefont {Scalapino}}, \bibinfo {author} {\bibfnamefont {R.~T.}\
  \bibnamefont {Scalettar}}, \ and\ \bibinfo {author} {\bibfnamefont {R.~L.}\
  \bibnamefont {Sugar}},\ }\href {\doibase 10.1103/PhysRevLett.74.2367}
  {\bibfield  {journal} {\bibinfo  {journal} {Phys. Rev. Lett.}\ }\textbf
  {\bibinfo {volume} {74}},\ \bibinfo {pages} {2367} (\bibinfo {year}
  {1995})}\BibitemShut {NoStop}%
\bibitem [{\citenamefont {Jarrell}(1995)}]{jarrell95}%
  \BibitemOpen
  \bibfield  {author} {\bibinfo {author} {\bibfnamefont {M.}~\bibnamefont
  {Jarrell}},\ }\href {\doibase 10.1103/PhysRevB.51.7429} {\bibfield  {journal}
  {\bibinfo  {journal} {Phys. Rev. B}\ }\textbf {\bibinfo {volume} {51}},\
  \bibinfo {pages} {7429} (\bibinfo {year} {1995})}\BibitemShut {NoStop}%
\bibitem [{\citenamefont {Rozenberg}(1995)}]{rozenberg95}%
  \BibitemOpen
  \bibfield  {author} {\bibinfo {author} {\bibfnamefont {M.~J.}\ \bibnamefont
  {Rozenberg}},\ }\href {\doibase 10.1103/PhysRevB.52.7369} {\bibfield
  {journal} {\bibinfo  {journal} {Phys. Rev. B}\ }\textbf {\bibinfo {volume}
  {52}},\ \bibinfo {pages} {7369} (\bibinfo {year} {1995})}\BibitemShut
  {NoStop}%
\bibitem [{\citenamefont {Georges}\ \emph {et~al.}(1996)\citenamefont
  {Georges}, \citenamefont {Kotliar}, \citenamefont {Krauth},\ and\
  \citenamefont {Rozenberg}}]{georges96}%
  \BibitemOpen
  \bibfield  {author} {\bibinfo {author} {\bibfnamefont {A.}~\bibnamefont
  {Georges}}, \bibinfo {author} {\bibfnamefont {G.}~\bibnamefont {Kotliar}},
  \bibinfo {author} {\bibfnamefont {W.}~\bibnamefont {Krauth}}, \ and\ \bibinfo
  {author} {\bibfnamefont {M.~J.}\ \bibnamefont {Rozenberg}},\ }\href {\doibase
  10.1103/RevModPhys.68.13} {\bibfield  {journal} {\bibinfo  {journal} {Rev.
  Mod. Phys.}\ }\textbf {\bibinfo {volume} {68}},\ \bibinfo {pages} {13}
  (\bibinfo {year} {1996})}\BibitemShut {NoStop}%
\bibitem [{\citenamefont {Tahvildar-Zadeh}\ \emph {et~al.}(1997)\citenamefont
  {Tahvildar-Zadeh}, \citenamefont {Jarrell},\ and\ \citenamefont
  {Freericks}}]{tahvildarzadeh97}%
  \BibitemOpen
  \bibfield  {author} {\bibinfo {author} {\bibfnamefont {A.~N.}\ \bibnamefont
  {Tahvildar-Zadeh}}, \bibinfo {author} {\bibfnamefont {M.}~\bibnamefont
  {Jarrell}}, \ and\ \bibinfo {author} {\bibfnamefont {J.~K.}\ \bibnamefont
  {Freericks}},\ }\href {\doibase 10.1103/PhysRevB.55.R3332} {\bibfield
  {journal} {\bibinfo  {journal} {Phys. Rev. B}\ }\textbf {\bibinfo {volume}
  {55}},\ \bibinfo {pages} {R3332} (\bibinfo {year} {1997})}\BibitemShut
  {NoStop}%
\bibitem [{\citenamefont {Vidhyadhiraja}\ \emph {et~al.}(2000)\citenamefont
  {Vidhyadhiraja}, \citenamefont {Tahvildar-Zadeh}, \citenamefont {Jarrell},\
  and\ \citenamefont {Krishnamurthy}}]{vidhyadhiraja00}%
  \BibitemOpen
  \bibfield  {author} {\bibinfo {author} {\bibfnamefont {N.~S.}\ \bibnamefont
  {Vidhyadhiraja}}, \bibinfo {author} {\bibfnamefont {A.~N.}\ \bibnamefont
  {Tahvildar-Zadeh}}, \bibinfo {author} {\bibfnamefont {M.}~\bibnamefont
  {Jarrell}}, \ and\ \bibinfo {author} {\bibfnamefont {H.~R.}\ \bibnamefont
  {Krishnamurthy}},\ }\href {http://stacks.iop.org/0295-5075/49/i=4/a=459}
  {\bibfield  {journal} {\bibinfo  {journal} {Europhys. Lett}\ }\textbf
  {\bibinfo {volume} {49}},\ \bibinfo {pages} {459} (\bibinfo {year}
  {2000})}\BibitemShut {NoStop}%
\bibitem [{\citenamefont {Pruschke}\ \emph {et~al.}(2000)\citenamefont
  {Pruschke}, \citenamefont {Bulla},\ and\ \citenamefont
  {Jarrell}}]{pruschke00}%
  \BibitemOpen
  \bibfield  {author} {\bibinfo {author} {\bibfnamefont {T.}~\bibnamefont
  {Pruschke}}, \bibinfo {author} {\bibfnamefont {R.}~\bibnamefont {Bulla}}, \
  and\ \bibinfo {author} {\bibfnamefont {M.}~\bibnamefont {Jarrell}},\ }\href
  {\doibase 10.1103/PhysRevB.61.12799} {\bibfield  {journal} {\bibinfo
  {journal} {Phys. Rev. B}\ }\textbf {\bibinfo {volume} {61}},\ \bibinfo
  {pages} {12799} (\bibinfo {year} {2000})}\BibitemShut {NoStop}%
\bibitem [{\citenamefont {Capponi}\ and\ \citenamefont
  {Assaad}(2001)}]{capponi01}%
  \BibitemOpen
  \bibfield  {author} {\bibinfo {author} {\bibfnamefont {S.}~\bibnamefont
  {Capponi}}\ and\ \bibinfo {author} {\bibfnamefont {F.~F.}\ \bibnamefont
  {Assaad}},\ }\href {\doibase 10.1103/PhysRevB.63.155114} {\bibfield
  {journal} {\bibinfo  {journal} {Phys. Rev. B}\ }\textbf {\bibinfo {volume}
  {63}},\ \bibinfo {pages} {155114} (\bibinfo {year} {2001})}\BibitemShut
  {NoStop}%
\bibitem [{\citenamefont {Benlagra}\ \emph {et~al.}(2011)\citenamefont
  {Benlagra}, \citenamefont {Pruschke},\ and\ \citenamefont
  {Vojta}}]{benlagra11}%
  \BibitemOpen
  \bibfield  {author} {\bibinfo {author} {\bibfnamefont {A.}~\bibnamefont
  {Benlagra}}, \bibinfo {author} {\bibfnamefont {T.}~\bibnamefont {Pruschke}},
  \ and\ \bibinfo {author} {\bibfnamefont {M.}~\bibnamefont {Vojta}},\ }\href
  {\doibase 10.1103/PhysRevB.84.195141} {\bibfield  {journal} {\bibinfo
  {journal} {Phys. Rev. B}\ }\textbf {\bibinfo {volume} {84}},\ \bibinfo
  {pages} {195141} (\bibinfo {year} {2011})}\BibitemShut {NoStop}%
\bibitem [{\citenamefont {Wu}\ and\ \citenamefont {Tremblay}(2015)}]{wu15}%
  \BibitemOpen
  \bibfield  {author} {\bibinfo {author} {\bibfnamefont {W.}~\bibnamefont
  {Wu}}\ and\ \bibinfo {author} {\bibfnamefont {A.-M.-S.}\ \bibnamefont
  {Tremblay}},\ }\href {\doibase 10.1103/PhysRevX.5.011019} {\bibfield
  {journal} {\bibinfo  {journal} {Phys. Rev. X}\ }\textbf {\bibinfo {volume}
  {5}},\ \bibinfo {pages} {011019} (\bibinfo {year} {2015})}\BibitemShut
  {NoStop}%
\bibitem [{\citenamefont {Aulbach}\ \emph
  {et~al.}(2015{\natexlab{a}})\citenamefont {Aulbach}, \citenamefont {Assaad},\
  and\ \citenamefont {Potthoff}}]{aulbach15}%
  \BibitemOpen
  \bibfield  {author} {\bibinfo {author} {\bibfnamefont {M.~W.}\ \bibnamefont
  {Aulbach}}, \bibinfo {author} {\bibfnamefont {F.~F.}\ \bibnamefont {Assaad}},
  \ and\ \bibinfo {author} {\bibfnamefont {M.}~\bibnamefont {Potthoff}},\
  }\href {\doibase 10.1103/PhysRevB.92.235131} {\bibfield  {journal} {\bibinfo
  {journal} {Phys. Rev. B}\ }\textbf {\bibinfo {volume} {92}},\ \bibinfo
  {pages} {235131} (\bibinfo {year} {2015}{\natexlab{a}})}\BibitemShut
  {NoStop}%
\bibitem [{\citenamefont {Hu}\ \emph {et~al.}(2017)\citenamefont {Hu},
  \citenamefont {Scalettar}, \citenamefont {Huang},\ and\ \citenamefont
  {Moritz}}]{hu17}%
  \BibitemOpen
  \bibfield  {author} {\bibinfo {author} {\bibfnamefont {W.}~\bibnamefont
  {Hu}}, \bibinfo {author} {\bibfnamefont {R.~T.}\ \bibnamefont {Scalettar}},
  \bibinfo {author} {\bibfnamefont {E.~W.}\ \bibnamefont {Huang}}, \ and\
  \bibinfo {author} {\bibfnamefont {B.}~\bibnamefont {Moritz}},\ }\href
  {\doibase 10.1103/PhysRevB.95.235122} {\bibfield  {journal} {\bibinfo
  {journal} {Phys. Rev. B}\ }\textbf {\bibinfo {volume} {95}},\ \bibinfo
  {pages} {235122} (\bibinfo {year} {2017})}\BibitemShut {NoStop}%
\bibitem [{\citenamefont {Sch\"afer}\ \emph {et~al.}(2018)\citenamefont
  {Sch\"afer}, \citenamefont {Katanin}, \citenamefont {Kitatani}, \citenamefont
  {Toschi},\ and\ \citenamefont {Held}}]{Schafer18}%
  \BibitemOpen
  \bibfield  {author} {\bibinfo {author} {\bibfnamefont {T.}~\bibnamefont
  {Sch\"afer}}, \bibinfo {author} {\bibfnamefont {A.~A.}\ \bibnamefont
  {Katanin}}, \bibinfo {author} {\bibfnamefont {M.}~\bibnamefont {Kitatani}},
  \bibinfo {author} {\bibfnamefont {A.}~\bibnamefont {Toschi}}, \ and\ \bibinfo
  {author} {\bibfnamefont {K.}~\bibnamefont {Held}},\ }\href
  {https://arxiv.org/abs/1812.03821} {\bibfield  {journal} {\bibinfo  {journal}
  {arXiv:1812.03821}\ } (\bibinfo {year} {2018})}\BibitemShut {NoStop}%
\bibitem [{\citenamefont {Meyer}\ and\ \citenamefont
  {Nolting}(2000{\natexlab{a}})}]{meyer00}%
  \BibitemOpen
  \bibfield  {author} {\bibinfo {author} {\bibfnamefont {D.}~\bibnamefont
  {Meyer}}\ and\ \bibinfo {author} {\bibfnamefont {W.}~\bibnamefont
  {Nolting}},\ }\href {\doibase 10.1103/PhysRevB.61.13465} {\bibfield
  {journal} {\bibinfo  {journal} {Phys. Rev. B}\ }\textbf {\bibinfo {volume}
  {61}},\ \bibinfo {pages} {13465} (\bibinfo {year}
  {2000}{\natexlab{a}})}\BibitemShut {NoStop}%
\bibitem [{\citenamefont {Ono}\ \emph {et~al.}(1991)\citenamefont {Ono},
  \citenamefont {Miura}, \citenamefont {Matsuura},\ and\ \citenamefont
  {Kuroda}}]{ono91}%
  \BibitemOpen
  \bibfield  {author} {\bibinfo {author} {\bibfnamefont {Y.}~\bibnamefont
  {Ono}}, \bibinfo {author} {\bibfnamefont {K.}~\bibnamefont {Miura}}, \bibinfo
  {author} {\bibfnamefont {T.}~\bibnamefont {Matsuura}}, \ and\ \bibinfo
  {author} {\bibfnamefont {Y.}~\bibnamefont {Kuroda}},\ }\href {\doibase
  https://doi.org/10.1016/0921-4534(91)90961-W} {\bibfield  {journal} {\bibinfo
   {journal} {Physica C}\ }\textbf {\bibinfo {volume} {185}},\ \bibinfo {pages}
  {1669} (\bibinfo {year} {1991})}\BibitemShut {NoStop}%
\bibitem [{\citenamefont {Nakatsuji}\ \emph {et~al.}(2004)\citenamefont
  {Nakatsuji}, \citenamefont {Pines},\ and\ \citenamefont
  {Fisk}}]{Nakatsuji04}%
  \BibitemOpen
  \bibfield  {author} {\bibinfo {author} {\bibfnamefont {S.}~\bibnamefont
  {Nakatsuji}}, \bibinfo {author} {\bibfnamefont {D.}~\bibnamefont {Pines}}, \
  and\ \bibinfo {author} {\bibfnamefont {Z.}~\bibnamefont {Fisk}},\ }\href
  {\doibase 10.1103/PhysRevLett.92.016401} {\bibfield  {journal} {\bibinfo
  {journal} {Phys. Rev. Lett.}\ }\textbf {\bibinfo {volume} {92}},\ \bibinfo
  {pages} {016401} (\bibinfo {year} {2004})}\BibitemShut {NoStop}%
\bibitem [{\citenamefont {Curro}\ \emph {et~al.}(2004)\citenamefont {Curro},
  \citenamefont {Young}, \citenamefont {Schmalian},\ and\ \citenamefont
  {Pines}}]{Curro04}%
  \BibitemOpen
  \bibfield  {author} {\bibinfo {author} {\bibfnamefont {N.~J.}\ \bibnamefont
  {Curro}}, \bibinfo {author} {\bibfnamefont {B.-L.}\ \bibnamefont {Young}},
  \bibinfo {author} {\bibfnamefont {J.}~\bibnamefont {Schmalian}}, \ and\
  \bibinfo {author} {\bibfnamefont {D.}~\bibnamefont {Pines}},\ }\href
  {\doibase 10.1103/PhysRevB.70.235117} {\bibfield  {journal} {\bibinfo
  {journal} {Phys. Rev. B}\ }\textbf {\bibinfo {volume} {70}},\ \bibinfo
  {pages} {235117} (\bibinfo {year} {2004})}\BibitemShut {NoStop}%
\bibitem [{\citenamefont {Yang}\ and\ \citenamefont {Pines}(2008)}]{Yang08}%
  \BibitemOpen
  \bibfield  {author} {\bibinfo {author} {\bibfnamefont {Y.-F.}\ \bibnamefont
  {Yang}}\ and\ \bibinfo {author} {\bibfnamefont {D.}~\bibnamefont {Pines}},\
  }\href {\doibase 10.1103/PhysRevLett.100.096404} {\bibfield  {journal}
  {\bibinfo  {journal} {Phys. Rev. Lett.}\ }\textbf {\bibinfo {volume} {100}},\
  \bibinfo {pages} {096404} (\bibinfo {year} {2008})}\BibitemShut {NoStop}%
\bibitem [{\citenamefont {Yang}\ \emph {et~al.}(2008)\citenamefont {Yang},
  \citenamefont {Fisk}, \citenamefont {Lee}, \citenamefont {Thompson},\ and\
  \citenamefont {Pines}}]{Yang08b}%
  \BibitemOpen
  \bibfield  {author} {\bibinfo {author} {\bibfnamefont {Y.-F.}\ \bibnamefont
  {Yang}}, \bibinfo {author} {\bibfnamefont {Z.}~\bibnamefont {Fisk}}, \bibinfo
  {author} {\bibfnamefont {H.-O.}\ \bibnamefont {Lee}}, \bibinfo {author}
  {\bibfnamefont {J.}~\bibnamefont {Thompson}}, \ and\ \bibinfo {author}
  {\bibfnamefont {D.}~\bibnamefont {Pines}},\ }\href {\doibase
  10.1038/nature07157} {\bibfield  {journal} {\bibinfo  {journal} {Nature
  (London)}\ }\textbf {\bibinfo {volume} {454}},\ \bibinfo {pages} {611}
  (\bibinfo {year} {2008})}\BibitemShut {NoStop}%
\bibitem [{\citenamefont {Shirer}\ \emph {et~al.}(2012)\citenamefont {Shirer},
  \citenamefont {Shockley}, \citenamefont {Dioguardi}, \citenamefont {Crocker},
  \citenamefont {Lin}, \citenamefont {Roberts-Warren}, \citenamefont {Nisson},
  \citenamefont {Klavins}, \citenamefont {Cooley}, \citenamefont {Yang},\ and\
  \citenamefont {Curro}}]{Shirer12}%
  \BibitemOpen
  \bibfield  {author} {\bibinfo {author} {\bibfnamefont {K.~R.}\ \bibnamefont
  {Shirer}}, \bibinfo {author} {\bibfnamefont {A.~C.}\ \bibnamefont
  {Shockley}}, \bibinfo {author} {\bibfnamefont {A.~P.}\ \bibnamefont
  {Dioguardi}}, \bibinfo {author} {\bibfnamefont {J.}~\bibnamefont {Crocker}},
  \bibinfo {author} {\bibfnamefont {C.~H.}\ \bibnamefont {Lin}}, \bibinfo
  {author} {\bibfnamefont {N.}~\bibnamefont {Roberts-Warren}}, \bibinfo
  {author} {\bibfnamefont {D.~M.}\ \bibnamefont {Nisson}}, \bibinfo {author}
  {\bibfnamefont {P.}~\bibnamefont {Klavins}}, \bibinfo {author} {\bibfnamefont
  {J.~C.}\ \bibnamefont {Cooley}}, \bibinfo {author} {\bibfnamefont {Y.-F.}\
  \bibnamefont {Yang}}, \ and\ \bibinfo {author} {\bibfnamefont {N.~J.}\
  \bibnamefont {Curro}},\ }\href {\doibase 10.1073/pnas.1209609109} {\bibfield
  {journal} {\bibinfo  {journal} {Proceedings of the National Academy of
  Sciences}\ }\textbf {\bibinfo {volume} {109}},\ \bibinfo {pages} {E3067}
  (\bibinfo {year} {2012})}\BibitemShut {NoStop}%
\bibitem [{\citenamefont {Wirth}\ and\ \citenamefont
  {Steglich}(2016)}]{wirth16}%
  \BibitemOpen
  \bibfield  {author} {\bibinfo {author} {\bibfnamefont {S.}~\bibnamefont
  {Wirth}}\ and\ \bibinfo {author} {\bibfnamefont {F.}~\bibnamefont
  {Steglich}},\ }\href {\doibase 10.1038/natrevmats.2016.51} {\bibfield
  {journal} {\bibinfo  {journal} {Nature Reviews Materials}\ }\textbf {\bibinfo
  {volume} {1}},\ \bibinfo {pages} {16051} (\bibinfo {year}
  {2016})}\BibitemShut {NoStop}%
\bibitem [{\citenamefont {Yang}\ \emph {et~al.}(2017)\citenamefont {Yang},
  \citenamefont {Pines},\ and\ \citenamefont {Lonzarich}}]{Yang17}%
  \BibitemOpen
  \bibfield  {author} {\bibinfo {author} {\bibfnamefont {Y.-F.}\ \bibnamefont
  {Yang}}, \bibinfo {author} {\bibfnamefont {D.}~\bibnamefont {Pines}}, \ and\
  \bibinfo {author} {\bibfnamefont {G.}~\bibnamefont {Lonzarich}},\ }\href
  {\doibase 10.1073/pnas.1703172114} {\bibfield  {journal} {\bibinfo  {journal}
  {Proceedings of the National Academy of Sciences}\ }\textbf {\bibinfo
  {volume} {114}},\ \bibinfo {pages} {6250} (\bibinfo {year}
  {2017})}\BibitemShut {NoStop}%
\bibitem [{\citenamefont {Jiang}\ \emph {et~al.}(2014)\citenamefont {Jiang},
  \citenamefont {Curro},\ and\ \citenamefont {Scalettar}}]{Jiang14}%
  \BibitemOpen
  \bibfield  {author} {\bibinfo {author} {\bibfnamefont {M.}~\bibnamefont
  {Jiang}}, \bibinfo {author} {\bibfnamefont {N.~J.}\ \bibnamefont {Curro}}, \
  and\ \bibinfo {author} {\bibfnamefont {R.~T.}\ \bibnamefont {Scalettar}},\
  }\href {\doibase 10.1103/PhysRevB.90.241109} {\bibfield  {journal} {\bibinfo
  {journal} {Phys. Rev. B}\ }\textbf {\bibinfo {volume} {90}},\ \bibinfo
  {pages} {241109(R)} (\bibinfo {year} {2014})}\BibitemShut {NoStop}%
\bibitem [{\citenamefont {Jiang}\ and\ \citenamefont {Yang}(2017)}]{Jiang17}%
  \BibitemOpen
  \bibfield  {author} {\bibinfo {author} {\bibfnamefont {M.}~\bibnamefont
  {Jiang}}\ and\ \bibinfo {author} {\bibfnamefont {Y.-F.}\ \bibnamefont
  {Yang}},\ }\href {\doibase 10.1103/PhysRevB.95.235160} {\bibfield  {journal}
  {\bibinfo  {journal} {Phys. Rev. B}\ }\textbf {\bibinfo {volume} {95}},\
  \bibinfo {pages} {235160} (\bibinfo {year} {2017})}\BibitemShut {NoStop}%
\bibitem [{\citenamefont {Costa}\ \emph {et~al.}(2019)\citenamefont {Costa},
  \citenamefont {Mendes-Santos}, \citenamefont {Paiva}, \citenamefont {Curro},
  \citenamefont {dos Santos},\ and\ \citenamefont {Scalettar}}]{Costa18}%
  \BibitemOpen
  \bibfield  {author} {\bibinfo {author} {\bibfnamefont {N.~C.}\ \bibnamefont
  {Costa}}, \bibinfo {author} {\bibfnamefont {T.}~\bibnamefont
  {Mendes-Santos}}, \bibinfo {author} {\bibfnamefont {T.}~\bibnamefont
  {Paiva}}, \bibinfo {author} {\bibfnamefont {N.~J.}\ \bibnamefont {Curro}},
  \bibinfo {author} {\bibfnamefont {R.~R.}\ \bibnamefont {dos Santos}}, \ and\
  \bibinfo {author} {\bibfnamefont {R.~T.}\ \bibnamefont {Scalettar}},\ }\href
  {\doibase 10.1103/PhysRevB.99.195116} {\bibfield  {journal} {\bibinfo
  {journal} {Phys. Rev. B}\ }\textbf {\bibinfo {volume} {99}},\ \bibinfo
  {pages} {195116} (\bibinfo {year} {2019})}\BibitemShut {NoStop}%
\bibitem [{\citenamefont {Nozieres}(1998)}]{nozieres98}%
  \BibitemOpen
  \bibfield  {author} {\bibinfo {author} {\bibfnamefont {P.}~\bibnamefont
  {Nozieres}},\ }\href {\doibase 10.1007/s100510050571} {\bibfield  {journal}
  {\bibinfo  {journal} {Eur. Phys. J. B}\ }\textbf {\bibinfo {volume} {6}},\
  \bibinfo {pages} {447} (\bibinfo {year} {1998})}\BibitemShut {NoStop}%
\bibitem [{\citenamefont {Kaul}\ and\ \citenamefont {Vojta}(2007)}]{Kaul07}%
  \BibitemOpen
  \bibfield  {author} {\bibinfo {author} {\bibfnamefont {R.~K.}\ \bibnamefont
  {Kaul}}\ and\ \bibinfo {author} {\bibfnamefont {M.}~\bibnamefont {Vojta}},\
  }\href {\doibase 10.1103/PhysRevB.75.132407} {\bibfield  {journal} {\bibinfo
  {journal} {Phys. Rev. B}\ }\textbf {\bibinfo {volume} {75}},\ \bibinfo
  {pages} {132407} (\bibinfo {year} {2007})}\BibitemShut {NoStop}%
\bibitem [{\citenamefont {Watanabe}\ and\ \citenamefont
  {Ogata}(2010)}]{Watanabe10}%
  \BibitemOpen
  \bibfield  {author} {\bibinfo {author} {\bibfnamefont {H.}~\bibnamefont
  {Watanabe}}\ and\ \bibinfo {author} {\bibfnamefont {M.}~\bibnamefont
  {Ogata}},\ }\href {\doibase 10.1103/PhysRevB.81.113111} {\bibfield  {journal}
  {\bibinfo  {journal} {Phys. Rev. B}\ }\textbf {\bibinfo {volume} {81}},\
  \bibinfo {pages} {113111} (\bibinfo {year} {2010})}\BibitemShut {NoStop}%
\bibitem [{\citenamefont {Burdin}\ and\ \citenamefont {Lacroix}()}]{Burdin18}%
  \BibitemOpen
  \bibfield  {author} {\bibinfo {author} {\bibfnamefont {S.}~\bibnamefont
  {Burdin}}\ and\ \bibinfo {author} {\bibfnamefont {C.}~\bibnamefont
  {Lacroix}},\ }\href {https://arxiv.org/abs/1810.05383} {\bibinfo  {journal}
  {arXiv:1810.05383}\ }\BibitemShut {NoStop}%
\bibitem [{\citenamefont {Blankenbecler}\ \emph {et~al.}(1981)\citenamefont
  {Blankenbecler}, \citenamefont {Scalapino},\ and\ \citenamefont
  {Sugar}}]{blankenbecler81}%
  \BibitemOpen
\bibfield  {journal} {  }\bibfield  {author} {\bibinfo {author} {\bibfnamefont
  {R.}~\bibnamefont {Blankenbecler}}, \bibinfo {author} {\bibfnamefont {D.~J.}\
  \bibnamefont {Scalapino}}, \ and\ \bibinfo {author} {\bibfnamefont {R.~L.}\
  \bibnamefont {Sugar}},\ }\href {\doibase 10.1103/PhysRevD.24.2278} {\bibfield
   {journal} {\bibinfo  {journal} {Phys. Rev. D}\ }\textbf {\bibinfo {volume}
  {24}},\ \bibinfo {pages} {2278} (\bibinfo {year} {1981})}\BibitemShut
  {NoStop}%
\bibitem [{\citenamefont {Hirsch}(1985)}]{hirsch85}%
  \BibitemOpen
  \bibfield  {author} {\bibinfo {author} {\bibfnamefont {J.~E.}\ \bibnamefont
  {Hirsch}},\ }\href {\doibase 10.1103/PhysRevB.31.4403} {\bibfield  {journal}
  {\bibinfo  {journal} {Phys. Rev. B}\ }\textbf {\bibinfo {volume} {31}},\
  \bibinfo {pages} {4403} (\bibinfo {year} {1985})}\BibitemShut {NoStop}%
\bibitem [{\citenamefont {White}\ \emph {et~al.}(1989)\citenamefont {White},
  \citenamefont {Scalapino}, \citenamefont {Sugar}, \citenamefont {Loh},
  \citenamefont {Gubernatis},\ and\ \citenamefont {Scalettar}}]{white89a}%
  \BibitemOpen
  \bibfield  {author} {\bibinfo {author} {\bibfnamefont {S.~R.}\ \bibnamefont
  {White}}, \bibinfo {author} {\bibfnamefont {D.~J.}\ \bibnamefont
  {Scalapino}}, \bibinfo {author} {\bibfnamefont {R.~L.}\ \bibnamefont
  {Sugar}}, \bibinfo {author} {\bibfnamefont {E.~Y.}\ \bibnamefont {Loh}},
  \bibinfo {author} {\bibfnamefont {J.~E.}\ \bibnamefont {Gubernatis}}, \ and\
  \bibinfo {author} {\bibfnamefont {R.~T.}\ \bibnamefont {Scalettar}},\ }\href
  {\doibase 10.1103/PhysRevB.40.506} {\bibfield  {journal} {\bibinfo  {journal}
  {Phys. Rev. B}\ }\textbf {\bibinfo {volume} {40}},\ \bibinfo {pages} {506}
  (\bibinfo {year} {1989})}\BibitemShut {NoStop}%
\bibitem [{\citenamefont {Assaad}(2002)}]{assaad02}%
  \BibitemOpen
  \bibfield  {author} {\bibinfo {author} {\bibfnamefont {F.}~\bibnamefont
  {Assaad}},\ }in\ \href {https://www.huy.dev/Assad-LectureNotes-2002.pdf}
  {\emph {\bibinfo {booktitle} {Quantum Simulations of Complex Many-Body
  Systems: From Theory to Algorithms}}},\ Vol.~\bibinfo {volume} {10},\
  \bibinfo {editor} {edited by\ \bibinfo {editor} {\bibfnamefont
  {J.}~\bibnamefont {Grotendorst}}, \bibinfo {editor} {\bibfnamefont
  {D.}~\bibnamefont {Marx}}, \ and\ \bibinfo {editor} {\bibfnamefont
  {A.}~\bibnamefont {Muramatsu}}}\ (\bibinfo  {publisher} {NIC Series},\
  \bibinfo {year} {2002})\ pp.\ \bibinfo {pages} {99--156}\BibitemShut
  {NoStop}%
\bibitem [{\citenamefont {\surname{dos Santos}}(2003)}]{dosSantos03b}%
  \BibitemOpen
  \bibfield  {author} {\bibinfo {author} {\bibfnamefont {R.~R.}\ \bibnamefont
  {\surname{dos Santos}}},\ }\href {\doibase 10.1590/S0103-97332003000100003}
  {\bibfield  {journal} {\bibinfo  {journal} {Brazilian Journal of Physics}\
  }\textbf {\bibinfo {volume} {33}},\ \bibinfo {pages} {36 } (\bibinfo {year}
  {2003})}\BibitemShut {NoStop}%
\bibitem [{\citenamefont {Gubernatis}\ \emph {et~al.}(2016)\citenamefont
  {Gubernatis}, \citenamefont {Kawashima},\ and\ \citenamefont
  {Werner}}]{gubernatis16}%
  \BibitemOpen
  \bibfield  {author} {\bibinfo {author} {\bibfnamefont {J.}~\bibnamefont
  {Gubernatis}}, \bibinfo {author} {\bibfnamefont {N.}~\bibnamefont
  {Kawashima}}, \ and\ \bibinfo {author} {\bibfnamefont {P.}~\bibnamefont
  {Werner}},\ }\href
  {https://www.cambridge.org/us/academic/subjects/physics/condensed-matter-physics-nanoscience-and-mesoscopic-physics/quantum-monte-carlo-methods-algorithms-lattice-models?format=HB&isbn=9781107006423}
  {\emph {\bibinfo {title} {Quantum Monte Carlo Methods: Algorithms for Lattice
  Models}}}\ (\bibinfo  {publisher} {Cambridge University Press},\ \bibinfo
  {year} {2016})\BibitemShut {NoStop}%
\bibitem [{\citenamefont {Loh}\ \emph {et~al.}(1990)\citenamefont {Loh},
  \citenamefont {Gubernatis}, \citenamefont {Scalettar}, \citenamefont {White},
  \citenamefont {Scalapino},\ and\ \citenamefont {Sugar}}]{loh90}%
  \BibitemOpen
  \bibfield  {author} {\bibinfo {author} {\bibfnamefont {E.~Y.}\ \bibnamefont
  {Loh}}, \bibinfo {author} {\bibfnamefont {J.~E.}\ \bibnamefont {Gubernatis}},
  \bibinfo {author} {\bibfnamefont {R.~T.}\ \bibnamefont {Scalettar}}, \bibinfo
  {author} {\bibfnamefont {S.~R.}\ \bibnamefont {White}}, \bibinfo {author}
  {\bibfnamefont {D.~J.}\ \bibnamefont {Scalapino}}, \ and\ \bibinfo {author}
  {\bibfnamefont {R.~L.}\ \bibnamefont {Sugar}},\ }\href {\doibase
  10.1103/PhysRevB.41.9301} {\bibfield  {journal} {\bibinfo  {journal} {Phys.
  Rev. B}\ }\textbf {\bibinfo {volume} {41}},\ \bibinfo {pages} {9301}
  (\bibinfo {year} {1990})}\BibitemShut {NoStop}%
\bibitem [{\citenamefont {Troyer}\ and\ \citenamefont
  {Wiese}(2005)}]{troyer05}%
  \BibitemOpen
  \bibfield  {author} {\bibinfo {author} {\bibfnamefont {M.}~\bibnamefont
  {Troyer}}\ and\ \bibinfo {author} {\bibfnamefont {U.-J.}\ \bibnamefont
  {Wiese}},\ }\href {\doibase 10.1103/PhysRevLett.94.170201} {\bibfield
  {journal} {\bibinfo  {journal} {Phys. Rev. Lett.}\ }\textbf {\bibinfo
  {volume} {94}},\ \bibinfo {pages} {170201} (\bibinfo {year}
  {2005})}\BibitemShut {NoStop}%
\bibitem [{\citenamefont {Wei}\ and\ \citenamefont
  {Yang}(2017)}]{wei2017doping}%
  \BibitemOpen
  \bibfield  {author} {\bibinfo {author} {\bibfnamefont {L.-y.}\ \bibnamefont
  {Wei}}\ and\ \bibinfo {author} {\bibfnamefont {Y.-f.}\ \bibnamefont {Yang}},\
  }\href {\doibase 10.1038/srep46089} {\bibfield  {journal} {\bibinfo
  {journal} {Sci. Rep.}\ }\textbf {\bibinfo {volume} {7}},\ \bibinfo {pages}
  {46089} (\bibinfo {year} {2017})}\BibitemShut {NoStop}%
\bibitem [{\citenamefont {Huscroft}\ \emph {et~al.}(1999)\citenamefont
  {Huscroft}, \citenamefont {McMahan},\ and\ \citenamefont
  {Scalettar}}]{huscroft99}%
  \BibitemOpen
  \bibfield  {author} {\bibinfo {author} {\bibfnamefont {C.}~\bibnamefont
  {Huscroft}}, \bibinfo {author} {\bibfnamefont {A.~K.}\ \bibnamefont
  {McMahan}}, \ and\ \bibinfo {author} {\bibfnamefont {R.~T.}\ \bibnamefont
  {Scalettar}},\ }\href {\doibase 10.1103/PhysRevLett.82.2342} {\bibfield
  {journal} {\bibinfo  {journal} {Phys. Rev. Lett.}\ }\textbf {\bibinfo
  {volume} {82}},\ \bibinfo {pages} {2342} (\bibinfo {year}
  {1999})}\BibitemShut {NoStop}%
\bibitem [{\citenamefont {Held}\ \emph {et~al.}(2000)\citenamefont {Held},
  \citenamefont {Huscroft}, \citenamefont {Scalettar},\ and\ \citenamefont
  {McMahan}}]{held00}%
  \BibitemOpen
  \bibfield  {author} {\bibinfo {author} {\bibfnamefont {K.}~\bibnamefont
  {Held}}, \bibinfo {author} {\bibfnamefont {C.}~\bibnamefont {Huscroft}},
  \bibinfo {author} {\bibfnamefont {R.~T.}\ \bibnamefont {Scalettar}}, \ and\
  \bibinfo {author} {\bibfnamefont {A.~K.}\ \bibnamefont {McMahan}},\ }\href
  {\doibase 10.1103/PhysRevLett.85.373} {\bibfield  {journal} {\bibinfo
  {journal} {Phys. Rev. Lett.}\ }\textbf {\bibinfo {volume} {85}},\ \bibinfo
  {pages} {373} (\bibinfo {year} {2000})}\BibitemShut {NoStop}%
\bibitem [{\citenamefont {Titvinidze}\ \emph {et~al.}(2014)\citenamefont
  {Titvinidze}, \citenamefont {Schwabe},\ and\ \citenamefont
  {Potthoff}}]{Titvinidze14}%
  \BibitemOpen
  \bibfield  {author} {\bibinfo {author} {\bibfnamefont {I.}~\bibnamefont
  {Titvinidze}}, \bibinfo {author} {\bibfnamefont {A.}~\bibnamefont {Schwabe}},
  \ and\ \bibinfo {author} {\bibfnamefont {M.}~\bibnamefont {Potthoff}},\
  }\href {\doibase 10.1103/PhysRevB.90.045112} {\bibfield  {journal} {\bibinfo
  {journal} {Phys. Rev. B}\ }\textbf {\bibinfo {volume} {90}},\ \bibinfo
  {pages} {045112} (\bibinfo {year} {2014})}\BibitemShut {NoStop}%
\bibitem [{\citenamefont {Titvinidze}\ \emph {et~al.}(2015)\citenamefont
  {Titvinidze}, \citenamefont {Schwabe},\ and\ \citenamefont
  {Potthoff}}]{Titvinidze15}%
  \BibitemOpen
  \bibfield  {author} {\bibinfo {author} {\bibfnamefont {I.}~\bibnamefont
  {Titvinidze}}, \bibinfo {author} {\bibfnamefont {A.}~\bibnamefont {Schwabe}},
  \ and\ \bibinfo {author} {\bibfnamefont {M.}~\bibnamefont {Potthoff}},\
  }\href {\doibase 10.1140/epjb/e2014-50772-1} {\bibfield  {journal} {\bibinfo
  {journal} {Eur. Phys. J. B}\ }\textbf {\bibinfo {volume} {88}},\ \bibinfo
  {pages} {9} (\bibinfo {year} {2015})}\BibitemShut {NoStop}%
\bibitem [{\citenamefont {Aulbach}\ \emph
  {et~al.}(2015{\natexlab{b}})\citenamefont {Aulbach}, \citenamefont
  {Titvinidze},\ and\ \citenamefont {Potthoff}}]{Aulbach15a}%
  \BibitemOpen
  \bibfield  {author} {\bibinfo {author} {\bibfnamefont {M.~W.}\ \bibnamefont
  {Aulbach}}, \bibinfo {author} {\bibfnamefont {I.}~\bibnamefont {Titvinidze}},
  \ and\ \bibinfo {author} {\bibfnamefont {M.}~\bibnamefont {Potthoff}},\
  }\href {\doibase 10.1103/PhysRevB.91.174420} {\bibfield  {journal} {\bibinfo
  {journal} {Phys. Rev. B}\ }\textbf {\bibinfo {volume} {91}},\ \bibinfo
  {pages} {174420} (\bibinfo {year} {2015}{\natexlab{b}})}\BibitemShut
  {NoStop}%
\bibitem [{\citenamefont {Benali}\ \emph {et~al.}(2016)\citenamefont {Benali},
  \citenamefont {Bai}, \citenamefont {Curro},\ and\ \citenamefont
  {Scalettar}}]{benali16}%
  \BibitemOpen
  \bibfield  {author} {\bibinfo {author} {\bibfnamefont {A.}~\bibnamefont
  {Benali}}, \bibinfo {author} {\bibfnamefont {Z.~J.}\ \bibnamefont {Bai}},
  \bibinfo {author} {\bibfnamefont {N.~J.}\ \bibnamefont {Curro}}, \ and\
  \bibinfo {author} {\bibfnamefont {R.~T.}\ \bibnamefont {Scalettar}},\ }\href
  {\doibase 10.1103/PhysRevB.94.085132} {\bibfield  {journal} {\bibinfo
  {journal} {Phys. Rev. B}\ }\textbf {\bibinfo {volume} {94}},\ \bibinfo
  {pages} {085132} (\bibinfo {year} {2016})}\BibitemShut {NoStop}%
\bibitem [{\citenamefont {Costa}\ \emph {et~al.}(2018)\citenamefont {Costa},
  \citenamefont {Ara\'ujo}, \citenamefont {Lima}, \citenamefont {Paiva},
  \citenamefont {dos Santos},\ and\ \citenamefont {Scalettar}}]{Costa18a}%
  \BibitemOpen
  \bibfield  {author} {\bibinfo {author} {\bibfnamefont {N.~C.}\ \bibnamefont
  {Costa}}, \bibinfo {author} {\bibfnamefont {M.~V.}\ \bibnamefont {Ara\'ujo}},
  \bibinfo {author} {\bibfnamefont {J.~P.}\ \bibnamefont {Lima}}, \bibinfo
  {author} {\bibfnamefont {T.}~\bibnamefont {Paiva}}, \bibinfo {author}
  {\bibfnamefont {R.~R.}\ \bibnamefont {dos Santos}}, \ and\ \bibinfo {author}
  {\bibfnamefont {R.~T.}\ \bibnamefont {Scalettar}},\ }\href {\doibase
  10.1103/PhysRevB.97.085123} {\bibfield  {journal} {\bibinfo  {journal} {Phys.
  Rev. B}\ }\textbf {\bibinfo {volume} {97}},\ \bibinfo {pages} {085123}
  (\bibinfo {year} {2018})}\BibitemShut {NoStop}%
\bibitem [{\citenamefont {Trotter}(1959)}]{trotter59}%
  \BibitemOpen
  \bibfield  {author} {\bibinfo {author} {\bibfnamefont {H.}~\bibnamefont
  {Trotter}},\ }\href {http://mathscinet.ams.org/mathscinet-getitem?mr=108732}
  {\bibfield  {journal} {\bibinfo  {journal} {Proc. Amer. Math. Soc.}\ }\textbf
  {\bibinfo {volume} {10}},\ \bibinfo {pages} {545} (\bibinfo {year}
  {1959})}\BibitemShut {NoStop}%
\bibitem [{\citenamefont {Suzuki}(1976)}]{suzuki76}%
  \BibitemOpen
  \bibfield  {author} {\bibinfo {author} {\bibfnamefont {M.}~\bibnamefont
  {Suzuki}},\ }\href {https://academic.oup.com/ptp/article/56/5/1454/1860476}
  {\bibfield  {journal} {\bibinfo  {journal} {Prog. Theor. Phys.}\ }\textbf
  {\bibinfo {volume} {56}},\ \bibinfo {pages} {1454} (\bibinfo {year}
  {1976})}\BibitemShut {NoStop}%
\bibitem [{\citenamefont {Fye}(1986)}]{fye86}%
  \BibitemOpen
  \bibfield  {author} {\bibinfo {author} {\bibfnamefont {R.~M.}\ \bibnamefont
  {Fye}},\ }\href {\doibase 10.1103/PhysRevB.33.6271} {\bibfield  {journal}
  {\bibinfo  {journal} {Phys. Rev. B}\ }\textbf {\bibinfo {volume} {33}},\
  \bibinfo {pages} {6271} (\bibinfo {year} {1986})}\BibitemShut {NoStop}%
\bibitem [{\citenamefont {Scalettar}\ \emph {et~al.}(1991)\citenamefont
  {Scalettar}, \citenamefont {Noack},\ and\ \citenamefont
  {Singh}}]{scalettar91}%
  \BibitemOpen
  \bibfield  {author} {\bibinfo {author} {\bibfnamefont {R.~T.}\ \bibnamefont
  {Scalettar}}, \bibinfo {author} {\bibfnamefont {R.~M.}\ \bibnamefont
  {Noack}}, \ and\ \bibinfo {author} {\bibfnamefont {R.~R.~P.}\ \bibnamefont
  {Singh}},\ }\href {\doibase 10.1103/PhysRevB.44.10502} {\bibfield  {journal}
  {\bibinfo  {journal} {Phys. Rev. B}\ }\textbf {\bibinfo {volume} {44}},\
  \bibinfo {pages} {10502} (\bibinfo {year} {1991})}\BibitemShut {NoStop}%
\bibitem [{\citenamefont {Mondaini}\ \emph {et~al.}(2012)\citenamefont
  {Mondaini}, \citenamefont {Bouadim}, \citenamefont {Paiva},\ and\
  \citenamefont {dos Santos}}]{Mondaini12}%
  \BibitemOpen
  \bibfield  {author} {\bibinfo {author} {\bibfnamefont {R.}~\bibnamefont
  {Mondaini}}, \bibinfo {author} {\bibfnamefont {K.}~\bibnamefont {Bouadim}},
  \bibinfo {author} {\bibfnamefont {T.}~\bibnamefont {Paiva}}, \ and\ \bibinfo
  {author} {\bibfnamefont {R.~R.}\ \bibnamefont {dos Santos}},\ }\href
  {\doibase 10.1103/PhysRevB.85.125127} {\bibfield  {journal} {\bibinfo
  {journal} {Phys. Rev. B}\ }\textbf {\bibinfo {volume} {85}},\ \bibinfo
  {pages} {125127} (\bibinfo {year} {2012})}\BibitemShut {NoStop}%
\bibitem [{\citenamefont {Iglovikov}\ \emph {et~al.}(2015)\citenamefont
  {Iglovikov}, \citenamefont {Khatami},\ and\ \citenamefont
  {Scalettar}}]{iglovikov15}%
  \BibitemOpen
  \bibfield  {author} {\bibinfo {author} {\bibfnamefont {V.~I.}\ \bibnamefont
  {Iglovikov}}, \bibinfo {author} {\bibfnamefont {E.}~\bibnamefont {Khatami}},
  \ and\ \bibinfo {author} {\bibfnamefont {R.~T.}\ \bibnamefont {Scalettar}},\
  }\href {\doibase 10.1103/PhysRevB.92.045110} {\bibfield  {journal} {\bibinfo
  {journal} {Phys. Rev. B}\ }\textbf {\bibinfo {volume} {92}},\ \bibinfo
  {pages} {045110} (\bibinfo {year} {2015})}\BibitemShut {NoStop}%
\bibitem [{\citenamefont {Mermin}\ and\ \citenamefont
  {Wagner}(1966)}]{Mermin66}%
  \BibitemOpen
  \bibfield  {author} {\bibinfo {author} {\bibfnamefont {N.~D.}\ \bibnamefont
  {Mermin}}\ and\ \bibinfo {author} {\bibfnamefont {H.}~\bibnamefont
  {Wagner}},\ }\href {\doibase 10.1103/PhysRevLett.17.1133} {\bibfield
  {journal} {\bibinfo  {journal} {Phys. Rev. Lett.}\ }\textbf {\bibinfo
  {volume} {17}},\ \bibinfo {pages} {1133} (\bibinfo {year}
  {1966})}\BibitemShut {NoStop}%
\bibitem [{\citenamefont {Huse}(1988)}]{huse88}%
  \BibitemOpen
  \bibfield  {author} {\bibinfo {author} {\bibfnamefont {D.~A.}\ \bibnamefont
  {Huse}},\ }\href {\doibase 10.1103/PhysRevB.37.2380} {\bibfield  {journal}
  {\bibinfo  {journal} {Phys. Rev. B}\ }\textbf {\bibinfo {volume} {37}},\
  \bibinfo {pages} {2380} (\bibinfo {year} {1988})}\BibitemShut {NoStop}%
\bibitem [{\citenamefont {Hirsch}\ and\ \citenamefont {Fye}(1986)}]{hirsch86}%
  \BibitemOpen
  \bibfield  {author} {\bibinfo {author} {\bibfnamefont {J.~E.}\ \bibnamefont
  {Hirsch}}\ and\ \bibinfo {author} {\bibfnamefont {R.~M.}\ \bibnamefont
  {Fye}},\ }\href {\doibase 10.1103/PhysRevLett.56.2521} {\bibfield  {journal}
  {\bibinfo  {journal} {Phys. Rev. Lett.}\ }\textbf {\bibinfo {volume} {56}},\
  \bibinfo {pages} {2521} (\bibinfo {year} {1986})}\BibitemShut {NoStop}%
\bibitem [{\citenamefont {Meyer}\ and\ \citenamefont
  {Nolting}(2000{\natexlab{b}})}]{meyer00b}%
  \BibitemOpen
  \bibfield  {author} {\bibinfo {author} {\bibfnamefont {D.}~\bibnamefont
  {Meyer}}\ and\ \bibinfo {author} {\bibfnamefont {W.}~\bibnamefont
  {Nolting}},\ }\href {\doibase 10.1007/s100510070023} {\bibfield  {journal}
  {\bibinfo  {journal} {Eur. Phys. J. B}\ }\textbf {\bibinfo {volume} {18}},\
  \bibinfo {pages} {385} (\bibinfo {year} {2000}{\natexlab{b}})}\BibitemShut
  {NoStop}%
\bibitem [{\citenamefont {Batista}\ \emph {et~al.}(2003)\citenamefont
  {Batista}, \citenamefont {Bon\ifmmode~\check{c}\else \v{c}\fi{}a},\ and\
  \citenamefont {Gubernatis}}]{batista03}%
  \BibitemOpen
  \bibfield  {author} {\bibinfo {author} {\bibfnamefont {C.~D.}\ \bibnamefont
  {Batista}}, \bibinfo {author} {\bibfnamefont {J.}~\bibnamefont
  {Bon\ifmmode~\check{c}\else \v{c}\fi{}a}}, \ and\ \bibinfo {author}
  {\bibfnamefont {J.~E.}\ \bibnamefont {Gubernatis}},\ }\href {\doibase
  10.1103/PhysRevB.68.214430} {\bibfield  {journal} {\bibinfo  {journal} {Phys.
  Rev. B}\ }\textbf {\bibinfo {volume} {68}},\ \bibinfo {pages} {214430}
  (\bibinfo {year} {2003})}\BibitemShut {NoStop}%
\end{thebibliography}%

\end{document}